\documentclass{optica-article}
\journal{opticajournal} 
\articletype{Research Article}
\usepackage{graphicx}
\usepackage{amsmath,psfrag,textcomp}
\usepackage{caption}
\usepackage{subcaption}
\usepackage{comment}
\usepackage[title]{appendix}
\usepackage{overpic}
\usepackage{multirow}
\definecolor{lgray1}{rgb}{0.91, 0.91, 0.91}
\usepackage[export]{adjustbox}
\usepackage{bm}
\usepackage[usestackEOL]{stackengine}

\usepackage{float}

\newcommand{\bracket}[1]{\left\langle #1\right\rangle}

\newcommand{\be}{\begin{equation}}
\newcommand{\ee}{\end{equation}}
\newcommand{\bd}{\begin{displaymath}}
\newcommand{\ed}{\end{displaymath}}

\newcommand{\beeq}[1] {\begin{equation}\begin{split}#1\end{split}\end{equation}}

\newif\ifcomments
\commentstrue 

\begin{document}

\title{Optimal calibration of optical tweezers with arbitrary integration time and sampling frequencies -- A general framework}

\author{Laura P\'erez-García\authormark{1}, Martin Selin\authormark{1}, Antonio Ciarlo\authormark{1,2}, Alessandro Magazz\`u\authormark{1},  Giuseppe Pesce\authormark{1,2}, Antonio Sasso\authormark{2}, Giovanni Volpe\authormark{1,\dag}, Isaac P\'erez Castillo\authormark{3,\ddag} and Alejandro V. Arzola\authormark{4,*}}

\address{\authormark{1}Department of Physics, University of Gothenburg, 41296 Gothenburg, Sweden\\
\authormark{2} Department of Physics E. Pancini, University of Naples Federico II, Complesso Universitario Monte Sant'Angelo, Via Cintia, I-80126, Naples, Italy\\
\authormark{3}Departamento de F\'isica, Universidad Autónoma Metropolitana-Iztapalapa, San Rafael Atlixco 186, Ciudad de México 09340, Mexico\\
\authormark{4}Departamento de Física Cuántica y Fotónica, Instituto de Física, Universidad Nacional Autónoma de México, C.P. 04510, Cd. de México, México}

\email{\authormark{\dag}giovanni.volpe@physics.gu.se,\authormark{\ddag}iperez@izt.uam.mx,\authormark{*} alejandro@fisica.unam.mx} 

\begin{abstract}
Optical tweezers (OT) have become an essential technique in several fields of physics, chemistry, and biology as precise micromanipulation tools and microscopic force transducers. Quantitative measurements require the accurate calibration of the trap stiffness of the optical trap and the diffusion constant of the optically trapped particle. This is typically done by statistical estimators constructed from the position signal of the particle, which is recorded by a digital camera or a quadrant photodiode. The finite integration time and sampling frequency of the detector need to be properly taken into account. Here, we present a general approach based on the joint probability density function of the sampled trajectory that corrects exactly the biases due to the detector's finite integration time and limited sampling frequency, providing theoretical formulas for the most widely employed calibration methods: equipartition, mean squared displacement, autocorrelation, power spectral density, and force reconstruction via maximum-likelihood-estimator analysis (FORMA). Our results, tested with experiments and Monte Carlo simulations, will permit users of OT to confidently estimate the trap stiffness and diffusion constant, extending their use to a broader set of experimental conditions.
\end{abstract}
 
\section{Introduction}
Optical tweezers (OT) are a key enabling technology in a wide range of fields including single-molecule biophysics, cell biology, colloidal studies, chemistry, statistical physics, transport phenomena, and even quantum physics \cite{jones_marago_volpe_2015, Gieseler21, Volpe21_Roadmap, bustamante2011revisiting, zemanek2019perspective, roca2017quantifying, gonzalez2021levitodynamics}. They have been used to manipulate microscopic objects with forces ranging from femtonewtons ($10^{-15}\,\rm{N}$) to piconewtons ($10^{-12}\,\rm{N}$) \cite{Gieseler21}. Furthermore, they have also been used to characterize the mechanical properties of the (possibly complex) fluids where these particles are immersed \cite{tassieri2015microrheology,robertson2018optical,Gieseler21}.

In first approximation, OT exert a force $F$ on the trapped particle that is proportional to the displacement $x$ of the particle from the center of the trap, i.e., $F= -\kappa x$, where $\kappa$ is the trap stiffness. There are various ways to experimentally determine $\kappa$. Most of these methods rely on the tracking of the position of the particle as a function of time, typically measured by a digital camera or a quadrant photodiode \cite{crocker1996methods, Pralle_1999, pesce2020optical, jones_marago_volpe_2015, neuman2004optical}. From the stochastic trajectory, different statistical estimators are used to estimate the trap stiffness of the optical trap and the diffusion constant of the optically trapped particle \cite{Gieseler21}. For instance, the equipartition (EP) method estimates $\kappa$ relying on the fact that, at thermodynamic equilibrium, the probability distribution of the particle position follows the Boltzmann distribution \cite{Gieseler21, jones_marago_volpe_2015,neuman2004optical}. Other methods can additionally estimante the diffusion constant by relying on the mean square displacement (MSD)\cite{jones_marago_volpe_2015}, the autocorrelation function (ACF) \cite{volpe2007brownian}, the power spectral density (PSD) \cite{berg2004power} and the force reconstruction via maximum likelihood estimator (FORMA) \cite{Gieseler21, Perez2018}. Generally speaking, the accurate and precise calibration of OT will depend on several parameters involved in the detection of the position of the particle, such as the signal-to-noise ratio (SNR), the sampling frequency or bandwidth, the integration time, and the length of the acquired data.

Typically, an OT setup includes a microscope digital camera or a quadrant photodiode (QPD) to measure the position of the optically trapped particle.
QPDs represent an excellent solution to calibrate OT when dealing with a single static optical trap; they work at very high sampling frequencies (on the order of hundreds of kilohertz) and short intergation times (on the order of microseconds) with high SNR. 
Nevertheless, when a larger field of view is needed (e.g., in holographic optical tweezers trapping multiple particles at the same time), digital cameras are commonly used to visualize the whole sample, permitting the user to select the region of interest and the particular particle(s) to be analyzed
\cite{jones_marago_volpe_2015, neuman2004optical}; however, standard microscopy cameras have the disadvantage that have low sampling frequencies (usually up to some kilohertz) and long integration times (up to milliseconds), which can affect the quality of the OT calibration with standard methods. 

Several works have discussed the effect of the integration time and sampling frequency on the calibration of OT 
\cite{Bogdan2018, Wong2006, Mortensen2021, Sharpe2007,  Savin2005, Savin2005a, Michalet2012, Loosemore2017, Berglund2010, vanderHorst2010, Gong2006, Lansdorp2012}. The EP method is arguably the simplest one and, often, the natural starting point for more complex analyses, with the limitation that it only provides the stiffness of the optical trap but not the diffusion constant of the particle\cite{Bogdan2018, Wong2006, Mortensen2021}; a finite integration time and a limited sampling frequency have been shown to introduce artifacts in the reconstructed trapping potentials \cite{Bogdan2018}. There has also been some work exploring the effects of the integration time on different geometries of the trapping potential and in cases of anisotropic diffusion, which may have applications in nuclear magnetic resonance imaging techniques \cite{Mortensen2021}. For MSD-based methods, the effect of the integration time has been discussed in the context of rheological studies measuring the viscoelastic properties of the liquid where the particles are immersed \cite{Bogdan2018, Savin2005a, Savin2005, Michalet2012, Loosemore2017, Mortensen2021, Berglund2010}, relevant for the study of processes intertwined with, e.g., cellular dynamics (such as cell growth, stem cell differentiation,  cell crawling, wound healing, protein regulation, cell malignancy, and even cell death \cite{Loosemore2017, vanderHorst2010}). When using the PSD method, the finite integration time and limited sampling frequency result in an effective low-pass filtering of the signal whose signature can be identified in the acquired power spectra \cite{Sharpe2007, Gong2006, Wong2006, Lansdorp2012}. 

Recently, there has been a growing interest in applying techniques of classical information theory to the calibration of OT. 
For instance, estimations using posterior distributions and maximum posterior estimators (e.g., FORMA with an uninformative prior \cite{Perez2018}) are two of the  recently developed alternatives, which, nonetheless, still consider restrictive  ideal conditions, such as high sampling frequency and negligible integration time \cite{Perez2018, Volpe21_Roadmap, Richly2013a, Pedersen2016}. Using maximum likelihood estimators, Refs.~\cite{Berglund2010,Lansdorp2012,Michalet2012} were able to incorporate into the mathematical analysis a limited sampling frequency and integration time and, then, use concepts from information theory, such as Fisher information \cite{Berglund2010} and the Cramer-Rao bound \cite{Michalet2012}, to ascertain the bona fide of their techniques. Other works have considered also a covariance-based estimator \cite{ Vestergaard2014} 
 
While many works and methodologies are available, a general and comprehensive approach to tackle the effects of the integration time, sampling rate and trajectory duration in  standard calibration methods is still lacking. Also, some of the works mentioned previously rely on some intermediate procedures to fit the model's functions, typically by means of non-linear fitting algorithms, with the disadvantage of being time consuming and potentially adding errors to the final estimators. 

Here, we develop a general framework from which we derive analytical solutions that incorporate sampling frequency and integration time in their descriptions for the most commonly used methods to estimate the stiffness and the diffusion constant in OT (namely EP, MSD, ACF, PSD, and FORMA). We provide generalized analytical solutions for these methods, enabling accurate and precise estimations, independent of the frame rate and of the integration time of the camera. This will offer a framework for future research to tackle problems where fast or real time estimations of the stiffness and diffusion are required, such as those dealing with non equilibrium thermodynamics or molecular biology.

This article is organized as follows: First, in Sec.~\ref{sec:1:problem}, we highlight the problems that arise due to a finite integration time and a limited sampling frequency on the estimates  derived from  the standard calibration methods (EP, MSD, ACF, PSD, and FORMA), which assume ideal conditions (i.e., high sampling frequency and negligible integration time). 
In Sec.~\ref{sec:2:framework}, we derive analytical expressions that incorporate arbitrary integration time and sampling frequency for all the methods, and we further use them to estimate the stiffness and diffusion. A summary of the methods in their standard and generalized forms, including practical considerations, is presented in Table~\ref{tab:methods}. Then, in Sec.~\ref{sec_experimental_calibration}, we compare the performance of these new formulas, to which we will refer as \emph{generalized formulas}, to estimate the stiffness and diffusion constant against standard ones in a broad range of experimental conditions, going from low sampling frequencies and large integration times to high sampling frequencies and short integration times. Finally, in Sec.~\ref{sec_trajectorylength}, we show the performance of the generalized formulas to estimate stiffness and diffusion constant when controlling total time and sample number. At the end of the article, in Secs.~\ref{sec_discussion} and \ref{sec_conclussions}, we discuss our results and give general conclusions and future research directions. 
 
\section{Problems due finite sampling frequency and integration time}
\label{sec:1:problem}

Let us start with a  reminder of the standard calibration methods to later dwell on the effects that a finite sampling frequency and integration time may have on them. We assume that the effect of the OT on a particle is well described by the Langevin equation modelling the motion of a Brownian particle in the overdamped regime subject to a Hookean restoring force. For simplicity, we also restrict ourselves to the one-dimensional case:
\begin{equation}\label{eq:langevin}
\frac{dx(t)}{dt}=-\frac{\kappa}{\gamma}  x(t)+\sqrt{2D} W(t)\,.
\end{equation}
The diffusion constant is $D=k_{\rm B}\,T/\gamma$, with $k_{\rm B}$ the Boltzmann constant, $T$ the absolute temperature, $\gamma=3\pi\nu d_{\rm p}$ the drag coefficient, $d_{\rm p}$ the diameter of the particle, $\nu$ the viscosity of the medium, and $W(t)$ a white noise with zero mean and Dirac-delta-correlated variance $\bracket{Wt)W(t')}=\delta(t-t')$ \cite{Risken1996}.

A formal solution to Eq.~\eqref{eq:langevin} reads
\beeq{
x(t)=x_0 e^{- t/\tau_{\text{ot}}}+\sqrt{2D} \int_0^{t} ds W(s) e^{-(t-s)/\tau_{\text{ot}}}\,,
\label{eq:exactsolution}
}
where $x_0$ is the initial position at time $t=0$ and $\tau_{\text{ot}}= \gamma/\kappa$ is the characteristic relaxation time of the particle in the optical trap.
From this theoretical model, one can easily derive exact formulas for physical quantities of interest that will depend on the two parameters $\kappa$  and $D$. To infer these parameters, one must contrast the theoretical formulas with sample estimators constructed from a dataset containing a time series of the particle's position measured experimentally. Let our dataset  be $\mathcal{D}\equiv\{x_n\}_{n=1}^{N_{\rm s}}$, where $x_n\equiv x(t_n)$ are the particle's positions taken at regular times $t_n=n\Delta t=n/f_{\rm s}$, for $n=1,2,...,N_{\rm s}$, and $f_{\rm s}=1/\Delta t$ denotes the sampling frequency. A summary of the implementation and relevant functions involved in these methods are (for further details see Refs.\cite{Gieseler21, jones_marago_volpe_2015}):
\begin{itemize}
\item\textbf{Potential and EP analyses}. At thermodynamic equilibrium, the probability of finding the particle around a position $x$ is given by the Boltzmann distribution:
\begin{equation}
\label{eq:potential}
\rho(x)=\rho_0 e^{-\frac{U(x)}{k_{\rm B} T}}\,,
\end{equation}
where  $\rho_0$ is a normalization factor. In this case, the statistical estimator is, fairly simply, the histogram of the probability of finding the particle at a given position constructed from the dataset $\mathcal{D}$. The experimental potential is estimated by means of the histogram $h(x_b)$ of the position, with $x_b$ the bins' coordinates,
\begin{equation}\label{eq:estpot}
U(x_b)=-k_B T \log(h(x_b))+constant.
\end{equation}
If we further assume that the potential is harmonic, 
\begin{equation}\label{eq:convpot}
U(x)=\frac{1}{2}\kappa (x-x_{\rm eq})^2\,,
\end{equation}
with $x_{\rm eq}$ the stable equilibrium position, we can fit this model function to the estimates in Eq.~\eqref{eq:estpot} and obtain $\kappa$ as a free parameter. Equivalently, by the equipartition theorem we know that
\beeq{
\kappa=\frac{k_{\rm B} T}{\bracket{(x-x_{\rm eq})^2}}\,.
}
This implies that a simple way to infer the value of $\kappa$ is to use the dataset $\mathcal{D}$ to construct an estimate of the position's variance $\bracket{(x-x_{\rm eq})^2}$. Its statistical estimator is the so-called sample variance, usually denoted as $s^2_{N_{\rm s}}$. Thus,
\begin{equation}
\label{eq:convep}
\kappa=\frac{k_B T}{s^2_{N_{\rm s}}}\,,
\end{equation} 
with $s^2_{N_{\rm s}}=\frac{1}{N_{\rm s}-1}\sum^{N_{\rm s}}_{n=1}( x_n-x_{\rm{eq}})^2$  and $x_{\rm{eq}}=\frac{1}{N_{\rm s}}\sum^{N_{\rm s}}_{n=1}x_n$. Since the potential and the EP formulas are essentially equivalent, very similar results are expected when the number of data is high enough; therefore, in the following, we will focus only in the EP method. 

\item\textbf{MSD analysis}. The MSD is theoretically defined as 
\begin{equation}
\label{eq:convmsd}
\text{MSD}(\tau)=\bracket{[x(t+\tau)-x(t)]^2}=\frac{2 k_{\rm B} T}{\kappa}\left(1-e^{-\tau/\tau_{{\rm ot}}}\right)\,,
\end{equation}
for $t$ larger than $\tau_{{\rm ot}}$. Given the dataset $\mathcal{D}$, its statistical estimator is given by  
\begin{equation}\label{eq:estmsd}
\text{MSD}(\tau_\ell)=\frac{1}{N_{\rm s}-\ell}\sum^{N_{\rm s}-\ell}_{n=1}(x_{n+\ell}-x_n)^2\,,
\end{equation}
with $x_n=x(n\Delta t)$ and $\tau_\ell=\ell\Delta t$. The theoretical formula and its estimator are then fitted to infer $\kappa$ and $\tau_{{\rm ot}}$ (from which $D$ can be straightforwardly obtained).

\item\textbf{ACF analysis}. The autocorrelation function is defined as:
\begin{equation}
\label{eq:convacf}
\text{ACF}(\tau)=\bracket{x(t+\tau)x(t)}=\frac{k_{\rm B} T}{\kappa} e^{-\tau/\tau_{{\rm ot}}}\,,
 \end{equation}
again for $t$ larger than the characteristic relaxation time. In this case, its estimator is given by
\begin{equation}\label{eq:estacf}
\text{ACF}(\tau_\ell)=\frac{1}{N_{\rm s}-\ell}\sum^{N_{\rm s}-\ell}_{n=1}(x_{n+\ell}x_n)\,,
\end{equation}
which upon comparison with its theoretical counterpart allows us to infer $\kappa$ and $D$ (through $\tau_{{\rm ot}}$).

\item\textbf {PSD analysis}. For a continuous and infinitely long signal (solution to Eq.~\eqref{eq:langevin}), the PSD is given by
\begin{equation}
 P(f)=\frac{1}{2\pi^2}\frac{D}{f_{{\rm c}}^2+f^2}\,,
 \label{eq:convpsd}
\end{equation}
which depends on $f_{\rm c}=\kappa/(2\pi\gamma)$ and $D$. If, however, the signal is taken at discrete time steps, the above formula must be replaced by \cite{Gieseler21}:
\begin{equation}
\label{eq:convpsdal}
P(f)=\frac{{\Delta x}^2/f_{\rm s}}{1+c^2-2c\cos(2\pi f/f_{\rm s})}\,,
\end{equation}
where $\Delta x=((1-c^2)D/2\pi f_{{\rm c}})^{1/2}$ and $c=e^{-2\pi f_{{\rm c}}/f_{\rm s}}$. The corresponding sampled estimator of the PSD is computed by means of the Fast Fourier Transform (FFT),
 \beeq{
	P(f_k)=\frac{\bracket{|\hat{x}(f_k)|^2}}{T_{\rm s}}
	=\frac{\Delta t^2}{T_{\rm s}}
	\bracket{|\rm{FFT}\{\{x_j\}^{N_{\rm s}}_{j=1}\}_k|^2} 
 \label{eq:convPSD0}
 }
with $f_k=k/T_{{\rm s}}$ for $k=1,\ldots, (N_{\rm s}-1)/2$. In this case, $\bracket{(\cdots)}$ stands for the expected value estimated by averaging many experimental replicas or by data compression (typically obtained by moving average). 

\item\textbf{FORMA analysis}. The values of $\kappa$ and $D$ are directly computed from the experimental dataset $\mathcal{D}$ by maximizing the likelihood of the linear model. The linear model comes about discretizing the Langevin equation to linear order, and rescaling the noise accordingly, yielding:
\beeq{\label{eq:convforma0}
\frac{x_{n+1}-x_{n}}{\Delta t}=-\frac{\kappa}{\gamma }x_{n}+\sqrt{\frac{2D}{\Delta t}}w_n
}
with $w_n$ Gaussian variables of zero mean and unit variance. From here, the model's likelihood is:
\beeq{
\mathcal{L}(\{x_n\}_{n=1}^{N_{\rm s}})=\frac{1}{\left(2D\Delta t\right)^{N_{\rm s}/2}}\exp\left[-\frac{1}{2}\sum_{n=1}^{N_{\rm s}-1}\left(\frac{x_{n+1}-\left(1-\frac{\Delta t}{\tau_{\rm ot}}\right)x_{n}}{\sqrt{2D \Delta t}}\right)^2\right]
}
maximizing with respect to the parameters we obtain the maximum likelihood estimators:
 \begin{equation}\label{eq:convforma}
	\frac{\kappa}{\gamma}=-\frac{\sum_{n=1}^{N_{\rm s}-1}x_{n}\frac{x_{n+1} -x_{n}}{\Delta t}}{\sum_{n=1}^{N_{\rm s}-1} [x_{n}]^2}\,,\quad\quad
	D=\frac{\Delta t}{2 (N_{\rm s}-1)}\sum_{n=1}^{N_{\rm s}-1}\left(\frac{x_{n+1}-{x_{n}}}{\Delta t}+\frac{\kappa}{\gamma}x_{n}\right)^2\,.
\end{equation}
\end{itemize}

\begin{figure}
\includegraphics[trim={0 0 0 0},clip, scale=0.2]{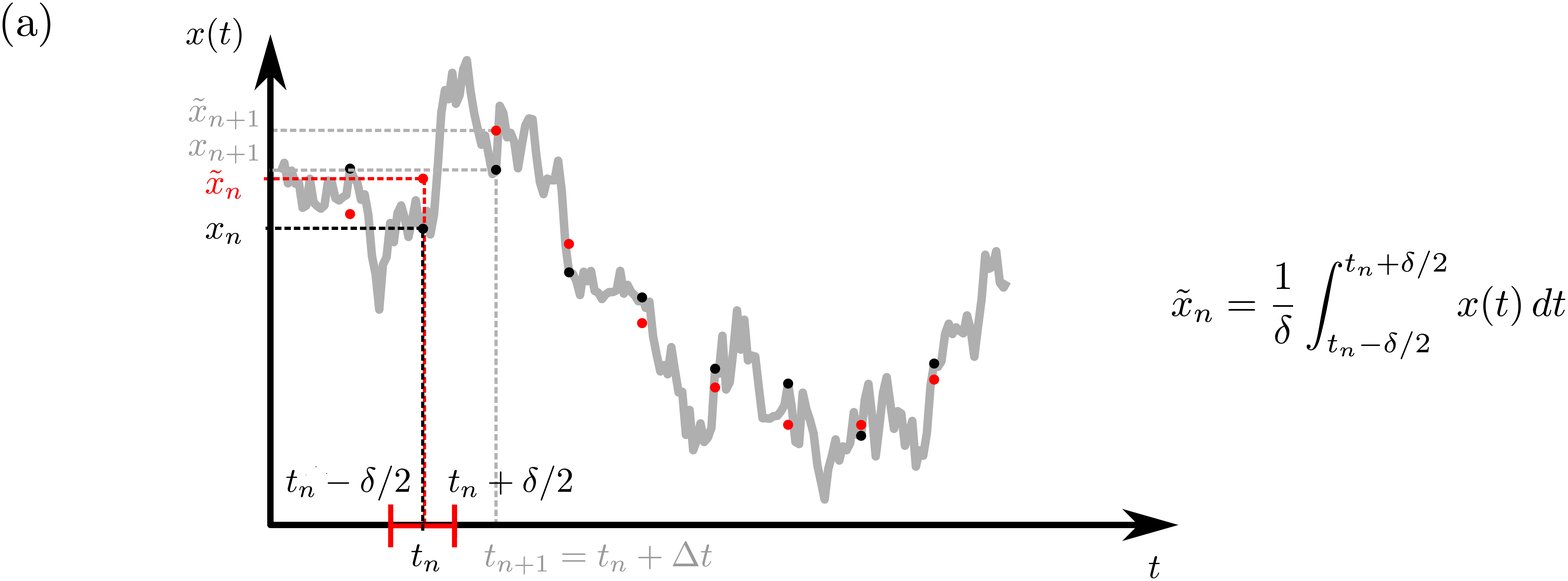}
\\
\includegraphics[trim={0 0 0 0},clip, scale=0.34]{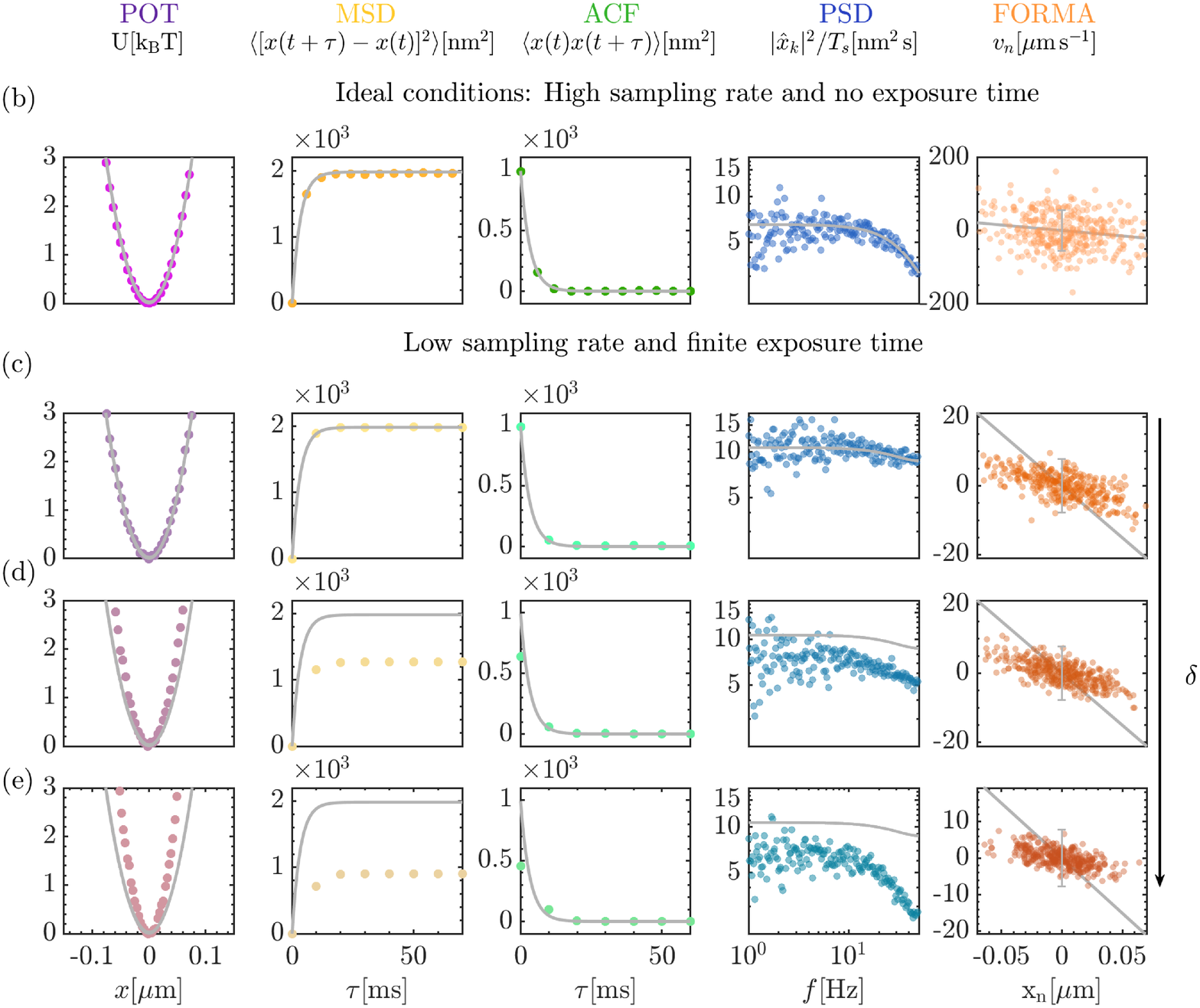}
\caption{{\bf Effect of low sampling frequency and long integration time on optical tweezers calibration.} 
(a) The trajectory of a particle (solid line) is sampled every $\Delta t$, instantaneously (black dots) or with a finite integration time  $\delta$ (red dots).  
(b-e) Behavior of the standard methods for various sampling frequencies and integration times on simulated trajectories generated through Monte Carlo simulations of a particle of diameter $d_{\rm p}=1.54\,\rm{\mu m}$ in a trap with stiffness $\kappa=4.08\,\rm{pN/\mu m}$ and diffusion constant $D= 0.299\,\rm{\mu m^2/s}$. 
The gray solid lines represent the analytical solutions of the standard methods, whereas the colored dotted lines represent their corresponding estimates obtained from the simulated data. From left to right, potential  (POT), mean square displacement (MSD), autocorrelation function (ACF), power spectrum density (PSD), and force reconstruction via maximum likelihood estimator (FORMA) methods.  
(b) When the conditions are ideal (i.e., with high sampling  frequency $f_{\rm s}=5000\,\rm{Hz}$ and short integration time $\delta=0\,\rm{s}$), there is good agreement between the estimators and the theoretical predictions (see Eqs.~\eqref{eq:convep}, \eqref{eq:convmsd}, \eqref{eq:convacf}, \eqref{eq:convpsdal}  and \eqref{eq:convforma}). 
(c-e) This agreement worsens as the conditions become less ideal (c) by lowering the sampling frequency to $f_{\rm s}= 100\,\rm{Hz}$ ($\delta=0\,\rm{m s}$), and then by increasing the integration time to (d) $\delta=5\,\rm{m s}$ and (e) $\delta=10\,\rm{m s}$.}
\label{fig:1:estimators_ideal_vs_notideal}
\end{figure}

All these methods assume that each sample contained in the dataset $\mathcal{D}$ is measured instantaneously. Additionally,  the MSD, ACF, PSD and FORMA methods need  the sampling frequency to be high enough in comparison with the characteristic time of the trap to fully access the time dependence of these observables. However, a real sampled trajectory will be distorted by the time the detector takes to collect the data, called \emph{integration time}, as schematically shown in Fig.~\ref{fig:1:estimators_ideal_vs_notideal}(a). This will skew the estimations of $\kappa$ and $D$. A way to model the effect of the integration time is to assume that the sampled position at a given time $t_n$ is averaged around a time window of size $\delta$, which represents the integration time, that is:
\begin{equation}
\label{eq:def}
\tilde{x}_n\equiv \frac{1}{\delta}\int_{t_n-\delta/2}^{t_n+\delta/2}  x(t)\, dt\,.
\end{equation}
The new  dataset $\tilde{\mathcal{D}}\equiv\{\tilde{x}_n\}_{n=1}^{N_{\rm s}}$ captures better the trajectory obtained in an experiments. This is equivalent to a low-pass filter, with an upper bound, $\delta\leq \Delta t$.
In a digital camera with global shutter, the integration time (exposure time) can be set in a range of values defined by the camera maker and is limited by the maximum sampling frequency (frame rate) allowed by the device. In a QPD, the effective integration time depends on the electronic bandwidth and typically is not a controllable parameter (in contrast to standard digital cameras, QPDs may reach quite easily sampling frequencies of the order of $\sim 10^5\,\rm{Hz}$, allowing very short integration times of the order of $10^{-5}\,\rm{s}$).

To appreciate the effect of the sampling frequency and the integration time on the standard methods previously described, we perform some Monte Carlo simulations of the dynamics of a particle trapped in an OT varying the sampling frequency and the integration time (see Refs.~\cite{jones_marago_volpe_2015} and Appendix~\ref{sec:simulations} for more details on the Monte Carlo simulations).  For this purpose, we used a particle of diameter $d_{\rm p}=1.54\,\rm{\mu m}$ and diffusion constant $D= 0.299 \,\rm{\mu m^2/s}$ trapped by an OT with $\kappa= 4.08 \,\rm{pN/\mu m}$. These values were chosen to be in line with the experiments discussed in Secs.~\ref{sec_experimental_calibration} and \ref{sec_trajectorylength}.

The scatter plots in Figs.~\ref{fig:1:estimators_ideal_vs_notideal}(b-e) show the behavior of the estimators under various acquisition conditions (from left to right, the methods based on the potential, MSD, ACF, PSD, and FORMA, estimated by means of Eqs.~\eqref{eq:estpot}, \eqref{eq:estmsd}, \eqref{eq:estacf}, and \eqref{eq:convPSD0}, respectively).  Figs.~\ref{fig:1:estimators_ideal_vs_notideal}(b) show the behavior for high sampling frequency, $f_{\rm s}=5000\,\rm{Hz}$, and zero integration time, while  Figs.~\ref{fig:1:estimators_ideal_vs_notideal}(c-e) show the behavior for low sampling frequency, $f_{\rm s}=100\,\rm{Hz}$, and three different integration times $\delta=0\,\rm{s}$, $5 \,\rm{ms}$, and $10\,\rm{ms}$ (the latter corresponds to the limit case where $\delta=\Delta t$). The solid gray lines show the behavior of the analytical expressions corresponding to the solutions of the standard methods considering the theoretical values of $\kappa$ and $D$ (Eqs.~\eqref{eq:convpot}, \eqref{eq:convmsd}, \eqref{eq:convacf}, \eqref{eq:convpsdal} and the deterministic part of Eq.~\eqref{eq:convforma0} for FORMA). 

Fig.~\ref{fig:1:estimators_ideal_vs_notideal}(b) shows the ideal case where the data points corresponding to each estimator follow the theoretical predictions of the standard formulas. In this case, the sampling frequency is more than one order of magnitude higher than the characteristic frequency of the trap, $f_{\rm s} = 5000\,\rm{Hz} \gg f_{\rm ot}=1/\tau_{\rm ot}=\kappa/\gamma=297\,\rm{Hz}$.

 A less ideal condition with still zero integration time but with a low sampling rate is shown in Figs.~\ref{fig:1:estimators_ideal_vs_notideal}(c). In this case, the sampling frequency, $f_{\rm s}=100\,\rm{Hz}$, is slightly lower than the characteristic frequency of the trap. Broadly speaking, we can already notice some important effects on the estimators that could have repercussions on the estimated values of $\kappa$ and $D$. Probably, the least affected is the potential, since this method does not depend on the sampling frequency (as the main condition to properly recover the potential or the variance of the position is that the total elapsed time of the trajectory is much larger than the relaxation time of the trap ($T_{\rm s} \gg \tau_{\rm ot}$) \cite{Gieseler21}, which is easily accomplished in the majority of cases since typically $\tau_{\rm ot}$ is of the order of some milliseconds and $T_{\rm s}$ could be on the order of seconds or more). 
 
 Regarding the MSD (second column in Figs.~\ref{fig:1:estimators_ideal_vs_notideal}(c)), the data of the estimator do not appear appreciably distorted. In this respect, the only condition needed to properly resolve the MSD with the estimator relies on a high enough sampling frequency  to recover the first part of the function. Since this curve reaches its maximum value at a characteristic time $\tau=\tau_{\rm ot}$, $\Delta t\lesssim\tau_{\rm ot}$ would enable enough data points in the first region of the plot. 
 
 The figure in the third column of Figs.~\ref{fig:1:estimators_ideal_vs_notideal}(c) shows the case of the ACF. The estimator in this case, similarly to what happens with the MSD, is not distorted by the low sampling frequency, following the standard formula. Nevertheless, since this function falls rapidly to zero with time, with a characteristic time given by $\tau_{\rm ot}$, from a statistical point of view, the behavior of the ACF would be better reproduced by the estimator if $\Delta t \lesssim\tau_{\rm ot}$, similarly to what happens with the MSD. 
 
 The estimator of the PSD shown in the fourth column of Figs.~\ref{fig:1:estimators_ideal_vs_notideal}(c) follows its corresponding analytical expression. Since the estimator of the PSD depends on the discrete Fourier transform, the aliasing of the data turns more evident as the sampling frequency gets lower, which is the case in this example. In this case, the analytical solution of the aliasing PSD is more adequate, which is in turn the one that is shown in these figures.  Under these still ideal conditions, the only restrictions for the PSD are to have a sufficiently long trajectory, with  $T_{\rm s}\gg\tau_{\rm ot}$, and high enough sampling frequency $f_{\rm s}\gtrsim f_{\rm ot}$ in order to have a good representation of the PSD at low and high frequencies around the corner frequency $f_{\rm c}$, which in our example is $f_{\rm c}=\kappa/(2 \pi \gamma)=47.3\,\rm{Hz}$. 
 
The last figure in Figs.~\ref{fig:1:estimators_ideal_vs_notideal}(c) depicts the behavior of FORMA. In this figure, in contrast to the one described above, where the sampling frequency was high, it is evident that the trend of the scattered plot, particularly its slope, is very different from the analytical formula. This was already expected since FORMA in its simple form was designed for data sets taken at sampling frequencies much higher than the characteristic frequencies of the trap \cite{Perez2018}, giving rise to wrong estimates of $\kappa$ and $D$ when these formulas were applied to experimental data under these conditions.
 
Figs.~\ref{fig:1:estimators_ideal_vs_notideal}(d-e) show the cases with low sampling frequency and a finite integration time, different from zero. These figures illustrate how the estimators of the different methods fail as the integration time increases.  In the case of the potential, as the integration time increases, the estimated values follow a steeper parabola than the expected one, which, according to Eq.~\eqref{eq:potential}, would indicate an apparent higher stiffness than the real one. In the same way, this effect would affect the estimation of the stiffness via the equipartition formula in its standard form (see Eq.~\eqref{eq:convep}). 

In these same conditions, the estimator of the MSD is distorted at all times, getting lower values than expected, with a lower initial slope than the theoretical prediction. According to the ideal formula (see Eq.~\eqref{eq:convmsd}), the MSD behaves linearly for short times with a slope of $2D$, while it reaches a plateau for long times at a value $2k_{\rm B}T/ \kappa$. Hence, comparing directly the estimated values of the MSD with a finite integration time and the corresponding standard formula (Eq.~\eqref{eq:convmsd}) would give rise to biased values of $D$ and $\kappa$, namely lower diffusion constant and higher stiffness than the expected ones. 

As the integration time increases, surprisingly, the ACF is not drastically distorted, as can be seen in the third column of Figs.~\ref{fig:1:estimators_ideal_vs_notideal}(d-e). The difference between the estimated values and the analytical standard description is quite subtle, albeit clearly visible in the data points evaluated at $\tau=0$. We will show in the following section that the zero lag time  of the ACF is actually the one that is mainly affected by the integration time while the following terms may present negligible distortions when $\delta$ has a low value. Considering this property, the ACF would enable more reliable estimations of $\kappa$ and $D$ than the other methods, particularly if the integration time is short or moderate and if the first data points of the ACF are ignored in the fitting procedure. 

For the case of the PSD, the integration time distorts the estimated PSD by reducing its expected values at short frequencies and increasing the slope of the decay at large frequencies. Since at short frequencies the dominant behavior of the PSD, according to the analytical expression Eq.~\eqref{eq:convpsdal}, is given by  $D/(2\pi) f_{\rm c}^{-2}$ and at large frequencies the PSD is expected to decay predominantly as $D/(2\pi) f^{-2}$ when the aliasing is ignored, similarly to the MSD, this give rise to an apparent increment of $\kappa$ and a reduction of $D$. 

In the case of FORMA, as the integration time increases, we see that the scatter plot becomes thinner and tilts counterclockwise, indicating again apparent higher and lower valued estimates of the stiffness and the diffusion constant, respectively.

\section{Generalized methods} 
\label{sec:2:framework}
As we have mentioned before, the dynamics of a particle of diameter $d_{\rm p}$, immersed in a fluid with viscosity $\nu$ and undergoning the action of an OT with stiffness $\kappa$, can be modelled by the Langevin equation~\eqref{eq:langevin}. Recall that the standard formulas of the methods, i.e., EP (Eq.~\eqref{eq:convpot} and \eqref{eq:convep}), MSD (Eq.~\eqref{eq:convmsd}), ACF (Eq.~\eqref{eq:convacf}), PSD (Eq.~\eqref{eq:convpsdal}) and FORMA (Eq.~\eqref{eq:convforma}) described above, do not account for the integration time, which implies that the inferred parameters will be biased. To incorporate this effect in a general framework, we should find the probability of observing a whole particle's trajectory already corrected for the integration time. If we were able to derive such an expression, then we could easily derive any physical quantity of interest from it. In the following, we show that this is indeed possible.

Starting from  Eq.~\eqref{eq:def}, using the formal solution Eq.~\eqref{eq:exactsolution}, and after some calculations based on the path integral formalism (see Appendix \ref{app:A}), it can be shown that the joint probability density function, denoted here as $p\left(\{\tilde{x}_n\}_{n=1}^{N_{\rm s}}\right)$ of observing the particle at sampled positions $\tilde{x}_1,\ldots, \tilde{x}_{N_{\rm s}}$ at times $0<t_1< \cdots <t_{N_{\rm s}}<T_{\rm s}$, is given by :
\beeq{
p(\{\tilde{x}_n\}_{n=1}^{N_{\rm s}})&=\frac{1}{\sqrt{(2\pi)^{N_{\rm s}} \det I}}\exp\left[-\frac{1}{2}\sum_{n,m=1}^{N_{\rm s}} \left(\tilde{x}_n- \tilde{x}_0 e^{- t_n/\tau_{\text{ot}}}\right)[I^{-1}]_{nm}\left(\tilde{x}_m- \tilde{x_0} e^{- t_m/\tau_{\text{ot}}}\right) \right]\,,
\label{eq:correctedjpdf}
}
where $\tilde{x}_0=\frac{\sinh(\alpha)}{\alpha} x_0$, and $[I^{-1}]_{nm}$ corresponds to the $(n,m)$-entry of the inverse covariance matrix between two  particle's blurry positions at time $t_n$ and $t_m$, 
\beeq{
I_{nm}&=\left[\frac{\sinh(\alpha)}{\alpha} \right]^2 D \tau_{\text{ot},x}\left[e^{-|t_n-t_m|/\tau_{\text{ot},x}}-e^{-( t_n+t_m)/\tau_{\text{ot},x}}\right]+D\tau_{\text{ot},x} \frac{\alpha-\cosh(\alpha)\sinh(\alpha)}{\alpha^2}\delta_{nm}\,,
\label{eq:correctedcovariance}
}
where $\delta_{nm}$ denotes the Kronecker delta and $\alpha=\delta/2\tau_{\text{ot}}$ is half the ratio between the integration time and the relaxation time. For the derivation of this formula, we have solely assumed that $|t_n-t_m|>\delta$, which is quite realistic. Besides, since we assume that the process runs within an interval starting at zero, we must have that the first time $t_1$ at which we have the first point recorded must obey $t_1-\delta>0$.

In the limit of $\alpha$ going to zero, which corresponds to either a very fast shutter velocity or a very long relaxation time, we recover the standard covariance matrix, and the joint PDF given by Eq.~\eqref{eq:correctedjpdf}, that can be written as $p(\{x_n\}_{n=1}^{N_{\rm s}})=p(x_0)\prod_{n=1}^{N_{\rm s}-1} G(x_{n+1}|x_n)$, with $ G(x_{n+1}|x_n)$ the propagator of the standard Ornstein–Uhlenbeck (OU) process.  Generally, the joint PDF  given by Eq. \eqref{eq:correctedjpdf} cannot be factorized in this way, since the corrected process is non-Markovian. Other than that, this expression contains all the statistical information of the system and we can now proceed to rederive the theoretical formulas used in the calibration methods. Henceforth, we will take the initial condition $x_0=0$ and, moreover, for MSD and ACF we will assume that enough time has elapsed so that we only keep the time-translational invariant term in the covariance matrix.

\begin{itemize}
\item\textbf{Potential and EP analysis}.
From Eq.~\eqref{eq:correctedjpdf}, we can evaluate the expectation value $\bracket{\tilde{x}^2_{N_{\rm s}}}$ which equals to $I_{nn}$. This automatically implies that
\beeq{\label{eq:revep}
\bracket{\tilde{x}^2_{N_{\rm s}}}=\frac{k_B T}{\kappa}\mathcal{F}(\alpha)\,\quad \text{with}\quad  \mathcal{F}(\alpha)=\frac{e^{-2\alpha}+2\alpha-1}{2\alpha^2}\,.
}
This implies, in turn, that the effective harmonic potential is given by the following formula: \begin{equation}\label{eq:revpot} 
    U(x)=\frac{1}{2}\frac{\kappa}{\mathcal{F}(\alpha)} x^2\,.
\end{equation}

When $\delta=0$, Eqs.~\eqref{eq:revep} and \eqref{eq:revpot} reduce to their standard form (Eqs.~\eqref{eq:convep} and \eqref{eq:convpot}). Notice that the formulas of the standard methods do not depend on $\tau_{\rm ot}$, so $\kappa$ can be directly determined from these formulas without any previous knowledge of the fluid properties. This must be contrasted to Eqs.~\eqref{eq:revep} and \eqref{eq:revpot}, which depend on $\tau_{\rm ot}$, through $\alpha$, and hence on the drag coefficient of the particle in the fluid, $\gamma$. In this sense, to directly use the generalized equations to predict the stiffness of the trap from a single experiment,  previous knowledge of $\gamma$ has to be incorporated in the solution. A different approach, as it was pointed out in Ref.~\cite{Wong2006}, would be to carry out some experiments, at least two, with different exposure times in order to obtain a data set of the variance $\langle\tilde{x}^2_{N_{\rm s}}\rangle(\delta)=k_B T\mathcal{F}(\delta/2\tau_{\rm ot})/\kappa$, and from these determine $\kappa$ and $\tau_{\rm ot}$ either by solving the resulting equations or by fitting the data points to the analytical formula. A similar approach can be followed in the potential method, with the difference that the variance  in this case is obtained through the fitting of the potential, Eq.~\eqref{eq:revpot}. For simplicity, in the following sections, we apply the EP using $\gamma=3\pi\nu d_{\rm p}$, where we assume that the viscosity, $\nu$, is the one of water at the laboratory temperature and $d_{\rm p}$ is the diameter of the particle reported by the fabricant.      

\item\textbf{ MSD analysis}.
The mean squared displacement for a lag time $\tau_\ell=t_{n+\ell}-t_{n}$ is defined as $\text{MSD}(\tau_\ell)=\bracket{(\tilde{x}_{n+\ell}-\tilde{x}_n)^2}$. Now, from the new joint PDF we have that
\beeq{
  \bracket{\tilde{x}_{n}^2}=\frac{k_B T}{\kappa}\mathcal{F}(\alpha)\,,\quad\bracket{\tilde{x}_n\tilde{x}_{n+\ell}}=\frac{k_B T}{\kappa}\left[\frac{\sinh(\alpha)}{\alpha} \right]^2 e^{-|\tau_{\ell}|/\tau_{\text{ot}}}\,,
}
and gathering these results yield
\beeq{\label{eq:revmsd}
\text{MSD}(\tau_\ell)&=\bracket{(\tilde{x}_{n+\ell}-\tilde{x}_n)^2}=\frac{2 k_B T}{\kappa}\left\{\mathcal{F}(\alpha)-\left[\frac{\sinh(\alpha)}{\alpha} \right]^2 e^{-|\tau_{\ell}|/\tau_{\text{ot}}}\right\}\,.
}

Fitting this equation to the experimental estimators allows us to infer the three free parameters $a=2 k_B T \mathcal{F}(\alpha)/\kappa$, $b=2 k_B T (\sinh(\alpha)/\alpha)^2/\kappa$, and $c=\tau_{\rm ot}$. Interestingly, we could infer $\tau_{\rm ot}$ and $\kappa$ from these three equations without knowing $\delta$ in advance, which actually could be inferred as well.  In our case, as we will see later, we use $\delta$ as a known parameter in all the realizations discussed below.     

\item\textbf{ ACF analysis}.
Recall that the autocorrelation function is defined as $\bracket{\tilde{x}_{m}\tilde{x}_n}=I_{mn}$, and therefore
\beeq{\label{eq:revacf}
\text{ACF}(t_n-t_m)=\bracket{\tilde{x}_m\tilde{x}_n}=\frac{k_B T}{\kappa}\left\{
\begin{array}{ll}
\left[\frac{\sinh(\alpha)}{\alpha} \right]^2  e^{-|t_n-t_m|/\tau_{\text{ot}}}& t_n\neq t_m\,,\\
\mathcal{F}(\alpha)& t_n= t_m\,.
\end{array}\right.
}
Interestingly, since $\mathcal{F}(\alpha)\leq \left[\frac{\sinh(\alpha)}{\alpha} \right]^2$ there is a jump from the first term at equal times to the following ones at different times in the autocorrelation function. Notice also that, for $\alpha$ small, the first leading correction to $\mathcal{F}(\alpha)$ is linear in $\alpha$, while it is quadratic for the factor $\left[\frac{\sinh(\alpha)}{\alpha} \right]^2$. Comparing the generalised equation Eq.~\eqref{eq:revacf} with the standard one, Eq.~\eqref{eq:convacf}, the terms corresponding to zero lag time ($t_n=t_m$) is the one that differs the most, while the behaviour for non-zero lag times has a more moderate correction to $\delta\neq0$. Overall, ignoring the zero-time term makes the ACF more resilient to trajectories sampled with a finite integration time. This explains the robustness of the standard ACF as the integration time increases (Fig.~\ref{fig:1:estimators_ideal_vs_notideal}(b-e)).

Looking at Eq.~\eqref{eq:revacf}, similarly to what happens to the MSD in its generalized form, one could divide the inference problems into two parts corresponding to zero and non-zero lag time and get the values of $\tau_{\rm ot}$, $\kappa$ and $\alpha$. For instance, we could use the non-zero lag time values to fit Eq.~\eqref{eq:revacf} and obtain the free parameters $a=k_B T (\sinh(\alpha)/\alpha)^2/\kappa$ and $b=\tau_{\rm ot}$ and the zero lag-time value to estimate $c=k_B T\mathcal{F}(\alpha)/\kappa$, then, we can compute $\kappa$, $\tau_{\rm ot}$ and $\alpha$, or equivalently $\delta$, from these three parameters. Importantly, this implies that we do not need to know $\delta$ in advance in order to infer $\kappa$ and $D$ accurately. In our case, as we will see later, we use $\delta$ as a known parameter in all the realizations discussed below. 

\item\textbf{ PSD analysis}.
Given the dataset $\tilde{\mathcal{D}}$ of the blurry trajectory, the PSD is defined  as follows:
\beeq{
P(f)=\frac{1}{N_{\rm s}\Delta t}\bracket{\left|\Delta t \sum_{n=1}^{N_{\rm s}} \tilde{x}_{n} e^{ 2\pi i f t_n}\right|^2}\,,
}
which can be rewritten as:
\beeq{
P(f)&=\frac{\Delta t}{N_{\rm s}}\sum_{n,m=1}^{N_{\rm s}}\bracket{\tilde{x}_n\tilde{x}_m}\left[\cos(2\pi f t_n)\cos(2\pi f t_m)+\sin(2\pi f t_n)\sin(2\pi f t_m)\right]\\
&=\frac{\Delta t}{N_{\rm s}}\sum_{n,m=1}^{N_{\rm s}} I_{nm}\cos[2\pi f( t_n- t_m)]\,,
}
where we have used the fact that $\bracket{\tilde{x}_n\tilde{x}_m}=I_{nm}$ is the covariance matrix. Gathering the previous results, we get the generalised formula for the PSD:
\beeq{
P(f)&=D\tau_{\text{ot}}\frac{\Delta t}{N_{\rm s}}\sum_{n,m=1}^{N_{\rm s}} \Bigg\{\left[\frac{\sinh(\alpha)}{\alpha} \right]^2 \left[e^{-|t_n-t_m|/\tau_{\text{ot}}}-e^{-( t_n+t_m)/\tau_{\text{ot}}}\right]\\
&+ \delta_{nm}\frac{\alpha-\cosh(\alpha)\sinh(\alpha)}{\alpha^2}\Bigg\}\cos[2\pi f( t_n- t_m)]\,.
}
After a bit of an unrelentingly tedious algebra, one shows that the double sum reads
\beeq{
&\sum_{n,m=1}^{N_{\rm s}} \left[e^{-|t_n-t_m|/\tau_{\text{ot}}}-e^{-( t_n+t_m)/\tau_{\text{ot}}}\right]\cos[2\pi f( t_n- t_m)]\\
&=N_{\rm s}\frac{\sinh(\Delta t/\tau_{\text{ot}})}{\cosh(\Delta t/\tau_{\text{ot}})-\cos(2\pi f\Delta t)}-e^{-(N_{\rm s}+1)\Delta t/\tau_{\text{ot}}}\frac{\cos(2\pi f N_{\rm s}\Delta t)-\cosh(N_{\rm s}\Delta t/\tau_{\text{ot}})}{\cos(2\pi f\Delta t)-\cosh(\Delta t/\tau_{\text{ot}})}\\
&-\frac{e^{-N_{\rm s}\Delta t/\tau_{\text{ot}}}}{\left(\cos(2\pi f\Delta t)-\cosh(\Delta t/\tau_{\text{ot}})\right)^2}\Bigg\{\cos(2\pi f N_{\rm s}\Delta t)\\
&+\sin(2\pi f\Delta t)\sin(2\pi f N_{\rm s}\Delta t)\sinh(\Delta t/\tau_{\text{ot}})-e^{N_{\rm s}\Delta t/\tau_{\text{ot}}}\\
&+\cos(2\pi f\Delta t)\cosh(\Delta t/\tau_{\text{ot}})\left[e^{N_{\rm s}\Delta t/\tau_{\text{ot}}}-\cos(2\pi f N_{\rm s}\Delta t)\right]\Bigg\}\,.
}
This allows us to get the final expression for the generalised  PSD formula:
\beeq{\label{eq:PSDcomplete}
P(f)&=D\tau_{\text{ot}}\frac{\Delta t}{N_{\rm s}}\left[\frac{\sinh(\alpha)}{\alpha} \right]^2\Bigg(N_{\rm s}\frac{\sinh(\Delta t/\tau_{\text{ot}})}{\cosh(\Delta t/\tau_{\text{ot}})-\cos(2\pi f\Delta t)}\\
&-e^{-(N_{\rm s}+1)\Delta t/\tau_{\text{ot}}}\frac{\cos(2\pi f N_{\rm s}\Delta t)-\cosh(N_{\rm s}\Delta t/\tau_{\text{ot}})}{\cos(2\pi f\Delta t)-\cosh(\Delta t/\tau_{\text{ot}})}\\
&-\frac{e^{-N_{\rm s}\Delta t/\tau_{\text{ot}}}}{\left(\cos(2\pi f\Delta t)-\cosh(\Delta t/\tau_{\text{ot}})\right)^2}\Bigg\{\cos(2\pi f N_{\rm s}\Delta t)\\
&+\sin(2\pi f\Delta t)\sin(2\pi f N_{\rm s}\Delta t)\sinh(\Delta t/\tau_{\text{ot}})-e^{N_{\rm s}\Delta t/\tau_{\text{ot}}}\\
&+\cos(2\pi f\Delta t)\cosh(\Delta t/\tau_{\text{ot}})\left[e^{N_{\rm s}\Delta t/\tau_{\text{ot}}}-\cos(2\pi f N_{\rm s}\Delta t)\right]\Bigg\}\Bigg)\\
&+D\tau_{\text{ot}}\Delta t \frac{\alpha-\cosh(\alpha)\sinh(\alpha)}{\alpha^2}\,.
}
Notice that the leading term for large $N_{\rm s}$ is:
\beeq{\label{eq:revpsdal}
P(f)&=D\tau_{\text{ot}}\Delta t\left(\left[\frac{\sinh(\alpha)}{\alpha} \right]^2\frac{\sinh(\Delta t/\tau_{\text{ot}})}{\cosh(\Delta t/\tau_{\text{ot}})-\cos(2\pi f\Delta t)}+\frac{\alpha-\cosh(\alpha)\sinh(\alpha)}{\alpha^2}\right)\\
&+\mathcal{O}(N_{\rm s}^{-1})\,.
}

\item\textbf{ Bayesian inference and maximum likelihood estimator (FORMA)}.
For the model at hand,  the likelihood $\mathcal{L}$ of observing the dataset $\tilde{\mathcal{D}}$ given the model's parameters, denoted here collectively as $\theta$, is:
\beeq{
\mathcal{L}(\tilde{\mathcal{D}}|\theta)=p(\{\tilde{x}_n\}_{n=1}^{N_{\rm s}}|\theta).
}
Using Bayes' rule, the posterior distribution of observing the parameters of the model given the dataset  is:
\beeq{
p(\theta|\{\tilde{x}_n\}_{n=1}^{N_{\rm s}})=\frac{\mathcal{L}(\tilde{\mathcal{D}}|\theta)p_0(\theta)}{\int d\theta \mathcal{L}(\tilde{\mathcal{D}}|\theta)p_0(\theta)}.
}
Unfortunately, the model's likelihood $\mathcal{L}(\tilde{\mathcal{D}}|\theta)$ is fairly complicated since the corrected process is non-Markovian, which implies that the inverse of the covariance matrix is not tridiagonal. Nevertheless, since $p(\{\tilde{x}_n\}_{n=1}^{N_{\rm s}}|\theta)$ is multivariate Gaussian, any marginal is multivariate Gaussian with the corresponding reduced covariance matrix. This allows us very easily to write the probability $G(\tilde{x}_{n+1})$ of observing the particle at the corrected position $\tilde{x}_{n+1}$ at at time $t_{n+1}$ conditioned that at a previous time $t_n$ it was at the corrected position $x_n$. Let us denote $\Delta t=t_{n+1}-t_n$. Then, one can show that
\beeq{
G(\tilde{x}_{n+1}|\tilde{x}_{n})=\exp\left[-\frac{1}{2\left(I_{n+1,n+1}-\frac{I^2_{n+1,n}}{I_{n,n}}\right)}\left(\tilde{x}_{n+1}- \frac{I_{n+1,n}}{I_{n,n}} \tilde{x}_n\right)^2\right],
}
but
\beeq{
I_{nn}&=I _{n+1,n+1}= \frac{k_BT}{\kappa}\mathcal{F}(\alpha)\,,\quad
I_{n+1,n}= \frac{k_BT}{\kappa}\left[\frac{\sinh(\alpha)}{\alpha} \right]^2 e^{-\Delta t/\tau_{\text{ot}}}\,,
}
and, therefore,
\begin{multline*}
G(\tilde{x}_{n+1}|\tilde{x}_{n})=\frac{1}{\sqrt{2 \pi\frac{k_BT}{\kappa_x}\mathcal{F}(\alpha)\left[1-\mathcal{G}^2(\alpha) e^{-2\Delta t/\tau_{\text{ot}}}\right]}}\times \\
\exp\left[-\frac{\left(\tilde{x}_{n+1}- \mathcal{G}(\alpha) e^{-\Delta t/\tau_{\text{ot}}} \tilde{x}_n\right)^2}{2\frac{k_BT}{\kappa}\mathcal{F}(\alpha)\left[1-\mathcal{G}^2(\alpha) e^{-2\Delta t/\tau_{\text{ot}}}\right]}\right]\,,
\end{multline*}

where we have introduced the function
\beeq{
\mathcal{G}(\alpha)=\frac{\left[\frac{\sinh(\alpha)}{\alpha} \right]^2}{\mathcal{F}(\alpha)}\,.
}
From here, we can approximate the model's likelihood as 
\beeq{
\mathcal{L}(\tilde{\mathcal{D}}|\theta)=\frac{\exp\left[-\sum_{n=1}^{N_{\rm s}}\frac{\left(\tilde{x}_{n+1}- \mathcal{G}(\theta_2\theta_3) e^{-\theta_2\Delta t} \tilde{x}_n\right)^2}{2\theta_1\mathcal{F}(\theta_2\theta_3)\left[1-\mathcal{G}^2(\theta_2\theta_3) e^{-2\theta_2\Delta t}\right]}\right]}{\left(2 \pi\theta_1\mathcal{F}(\theta_2\theta_3)\left[1-\mathcal{G}^2(\theta_2\theta_3) e^{-2\theta_2\Delta t}\right]\right)^{N_{\rm s}/2}}\,,
\label{eq:likelihood}
}
where we have considered as parameters $\theta=(\theta_1,\theta_2,\theta_3)=(\frac{k_BT}{\kappa},1/\tau_{\text{ot}},\delta/2)$. Let us next introduce the estimators:
\beeq{
\mathcal{T}_1=\frac{1}{N_{\rm s}}\sum_{n=1}^{N_{\rm s}}[\tilde{x}_{n+1}]^2\,,\quad \mathcal{T}_2=\frac{1}{N_{\rm s}}\sum_{n=1}^{N_{\rm s}}\tilde{x}_{n+1}\tilde{x}_{n}
\,,\quad \mathcal{T}_3=\frac{1}{N_{\rm s}}\sum_{n=1}^{N_{\rm s}}[\tilde{x}_{n}]^2\,.
}
Then, maximizing with respect to $\theta_1$ and $\theta_2$ we obtain the MLEs
\beeq{\label{eq:revforma}
\mathcal{G}(\theta^\star_2\theta_3) e^{-\theta^\star_2\Delta t}&=\frac{\mathcal{T}_2}{\mathcal{T}_3}\,,\quad\quad 
\theta_1^\star\mathcal{F}(\theta^\star_2\theta_3)=\frac{\mathcal{T}_1-\frac{\mathcal{T}_2^2}{\mathcal{T}_3}}{1-\left(\frac{\mathcal{T}_2}{\mathcal{T}_3}\right)^2}.
}
If we want to plot the dataset $\mathcal{D}$ in a two-dimensional scatter plot where the $x$-axis represents $\tilde{x}_{n}$ and  the $y$-axis represents $(\tilde{x}_n-\tilde{x}_n)/\Delta t$, then, looking at the argument of the exponential in Eq.~\eqref{eq:likelihood}, the cloud of experimental points will have an elliptical form with the major axis having a slope given by $-\left(1-\mathcal{T}_2/\mathcal{T}_3\right)/\Delta t$ and a width proportional to  $\sqrt{\mathcal{T}_1-\mathcal{T}_2^2/\mathcal{T}_3}/\Delta t$.
\end{itemize}
In the following section, we show the performance of these generalized solutions and compare them to the standard ones on experimental measurements.

\section{Experimental estimation of the stiffness and diffusion}
\label{sec_experimental_calibration}

To demonstrate the performance of the generalized formulas presented in the previous section, we performed a series of experiments with an OT. The details of the experimental setup are described in Appendix~\ref{sec:experiments}. 
Briefly, a silica particle of diameter $d_{\rm p}=1.54\pm 0.10\,\rm{\mu m}$ was trapped in water with an OT obtained by focusing a laser beam of $532\,\rm{nm}$ wavelength. The laboratory temperature was $T=22\,\pm0.5^\circ\rm{C}$. The trapped particle was recorded at three different sampling frequencies and four integration times (exposure time of the camera) for each frequency: $f_{\rm s}=500\,\rm{Hz}$  with $\delta=59$, $500$, $1000$, and $2000\,\mu \textrm{s}$; $f_{\rm s}=1499.25\,\rm{Hz}$ ($1500\,\rm{Hz}$ nominal value) with $\delta=59$, $200$, $350$ and $500\,\mu \textrm{s}$ and $f_{\rm s}=3496.5\,\rm{Hz}$  ($3500\,\rm{Hz}$ nominal value) with $\delta=59, 100, 150$ and $200\,\mu \textrm{s}$. In all these experiments, we acquired trajectories with $N_{\rm s}=10^5$ samples and in all the cases where fitting is required (e.g., MSD, ACF and PSD), we used standard non-linear least square fitting routines provided by MatLab.

\begin{figure}[H]
\includegraphics[trim={0 0 0 0},clip, scale=0.37]{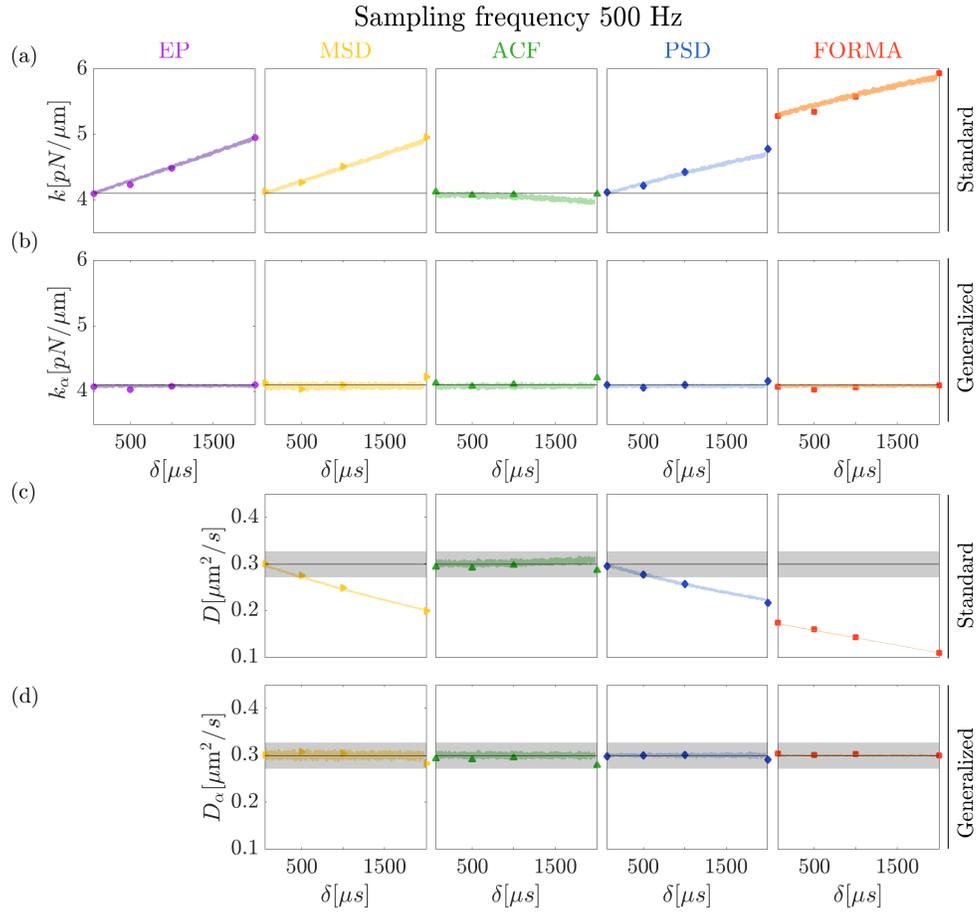}
\caption{{\bf Influence of $\delta$ on the estimation of the stiffness and diffusion: low sampling frequency.} (a,c) standard and (b,d) generalized formulas  at sampling frequency  $f_{\rm s}=500\,\rm{Hz}$. The data points show estimates from experimental realizations and the colored shaded areas depict confidence intervals obtained from simulations using $\bar{\kappa}_\alpha = 4.08\,\rm{pN/\mu m}$ and $\bar{D}_\alpha= 0.299\,\rm{\mu m^2/s}$.  As a reference, the gray shaded areas in (c,d) depict the range of the expected value $D^*=0.295\pm0.020\,\rm{\mu m^2/s}$. All the experimental data points where estimated using $N_{\rm s}=10^5$ samples. }
\label{fig:2:K_D_vs_Texpo_500hz}
\end{figure}

\begin{figure}[H]
\includegraphics[trim={0 0 0 0},clip, scale=0.37]{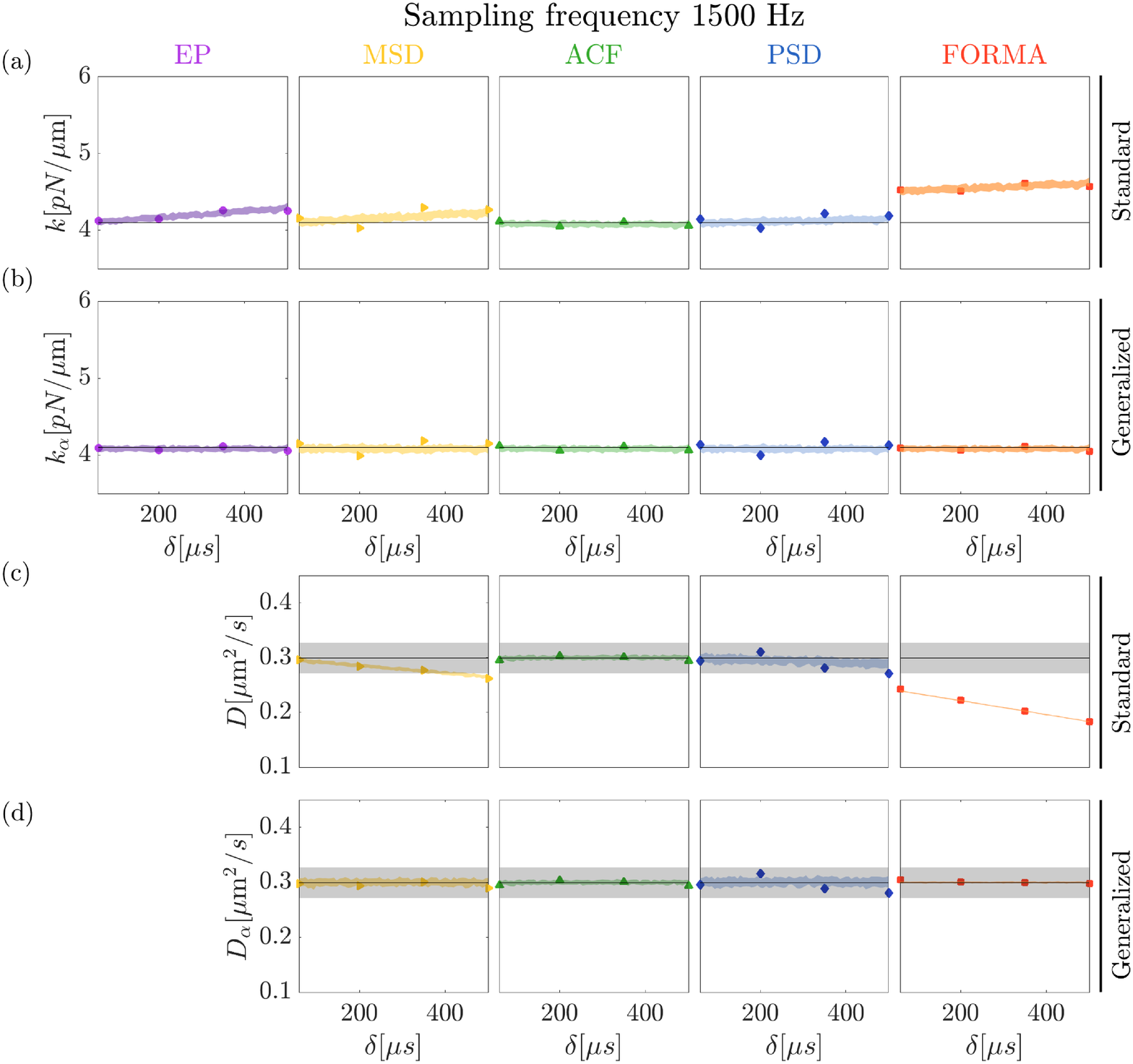}
\caption{{\bf Influence of $\delta$ on the estimation of the stiffness and diffusion: medium sampling frequency.} (a,c) standard and (b,d) generalized formulas at sampling frequency  $f_{\rm s}=1500\,\rm{Hz}$. The data points show estimates from experimental realizations and the colored shaded areas depict confidence intervals obtained from simulations using $\bar{\kappa}_\alpha = 4.08\,\rm{pN/\mu m}$ and $\bar{D}_\alpha= 0.299\,\rm{\mu m^2/s}$.  As a reference, the gray shaded areas in (c,d) depict the range of the expected value $D^*=0.295\pm0.020\,\rm{\mu m^2/s}$. All the experimental data points where estimated using $N_{\rm s}=10^5$ samples.}
\label{fig:3:K_D_vs_Texpo_1500hz}
\end{figure}

\begin{figure}[H]
\includegraphics[trim={0 0 0 0},clip, scale=0.37]{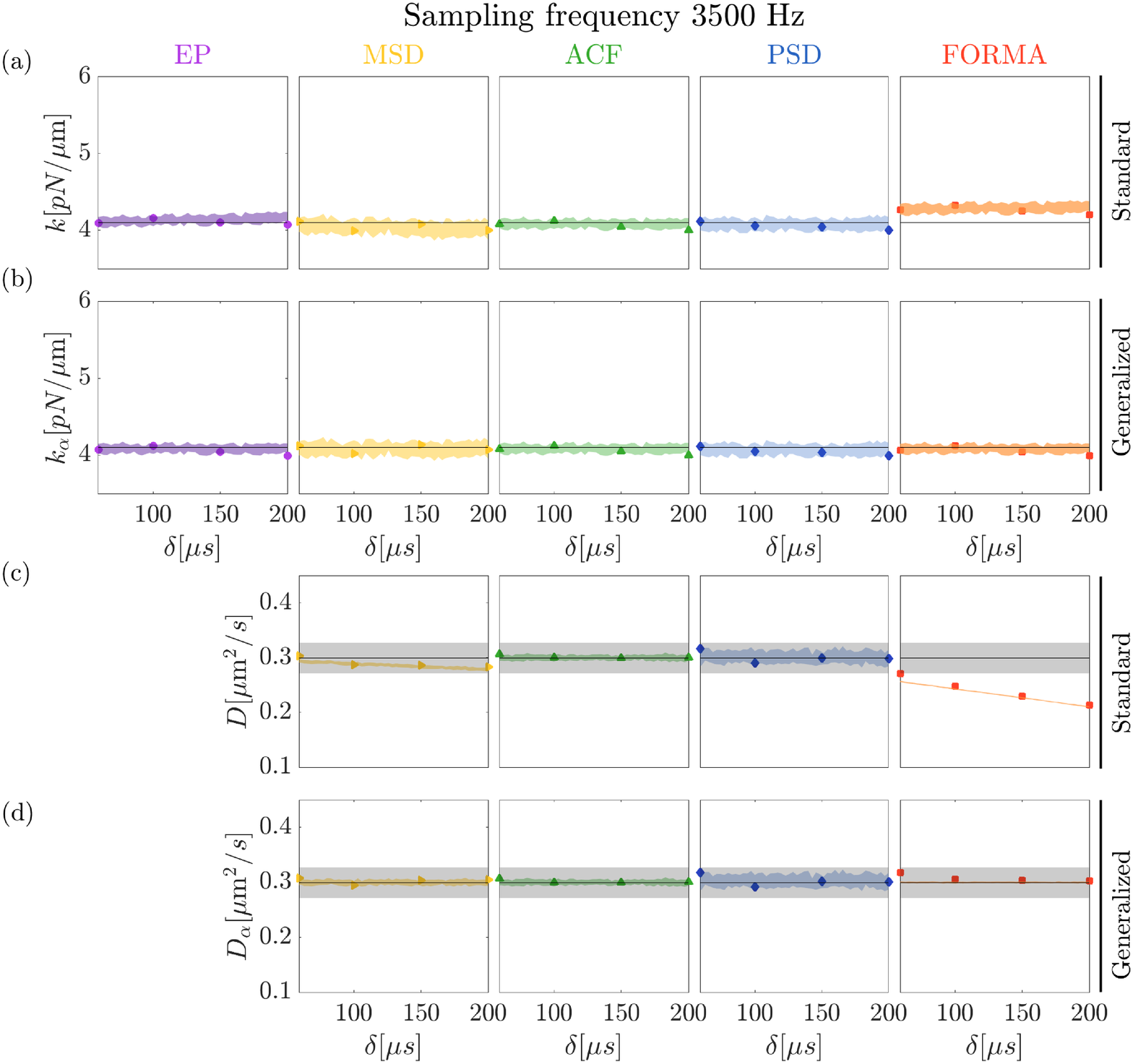}
\caption{{\bf Influence of $\delta$ on the estimation of the stiffness and diffusion: high sampling frequency.} (a,c) standard and (b,d) generalized formulas at sampling frequency $f_{\rm s}=3500\,\rm{Hz}$. The data points show estimates from experimental realizations and the colored shaded areas depict confidence intervals obtained from simulations using $\bar{\kappa}_\alpha = 4.08\,\rm{pN/\mu m}$ and $\bar{D}_\alpha= 0.299\,\rm{\mu m^2/s}$.  As a reference, the gray shaded areas in (c-d) depict the range of the expected value $D^*=0.295\pm0.020\,\rm{\mu m^2/s}$. All the experimental data points where estimated using $N_{\rm s}=10^5$ samples. }
\label{fig:4:K_D_vs_Texpo_3500hz}
\end{figure}

The estimated values of $\kappa$ and $D$ are shown in Figs.~\ref{fig:2:K_D_vs_Texpo_500hz}, \ref{fig:3:K_D_vs_Texpo_1500hz}, and \ref{fig:4:K_D_vs_Texpo_3500hz} for the three sampling frequencies $f_{\rm s}=500$, $1500$ and $3500\,\rm{Hz}$.  For each of these experiments, the values of $\kappa$ and $D$ were estimated as a function of $\delta$ following standard and generalized formulas. In Figs.~\ref{fig:2:K_D_vs_Texpo_500hz}, \ref{fig:3:K_D_vs_Texpo_1500hz}, and \ref{fig:4:K_D_vs_Texpo_3500hz}, the dots represent the experimental results. To compare our results with theory, we performed Monte Carlo simulations of the OU proccess (Eq.\eqref{eq:langevin}) taking the mean of the experimental values of $\kappa$ and $D$ retrieved by means of the generalized analyses, giving $\bar{\kappa}_\alpha=4.08\pm 0.05\,\rm{pN/\mu m}$  and $\bar{D}_\alpha= 0.299\pm 0.008\,\rm{\mu m^2/s}$. Performing 50 realizations with $N_{\rm s}=10^5$ samples for each case, the Monte Carlo simulations enable to estimate the confidence intervals obtained by means of the standard deviation of all the replicas as a function of $\delta$, which are depicted in colored shaded areas in Figs.~\ref{fig:2:K_D_vs_Texpo_500hz}, \ref{fig:3:K_D_vs_Texpo_1500hz}, and \ref{fig:4:K_D_vs_Texpo_3500hz}.  The gray shaded areas of the diffusion constant depict the  confidence interval of the expected diffusion computed from the size of the particle and the viscosity of water at the laboratory temperature  ($\nu=0.95\pm 0.01\,\rm{mPa\,s}$, with its error estimated from the error range of $T$), $D^*=0.295\pm0.020\,\rm{\mu m^2/s}$, which is useful as a reference parameter to evaluate our estimations. Additionally, using $\bar{\kappa}_\alpha$ and $\bar{D}_\alpha$, together with the Stokes-Einstein relation $D=k_B T/\gamma$, we estimate the characteristic time of our OT to be $\tau_{\rm{ot}}=\gamma/\kappa =3.4\,\rm{ms}$, which corresponds to the characteristic frequency $f_{\rm{ot}}= 1/\tau_{\rm{ot}}=297\,\rm{Hz}$.

As already illustrated in Fig.~\ref{fig:1:estimators_ideal_vs_notideal}, the standard formulas fail to give a good description of the main quantities as $\delta$ increases. This gives rise to wrong estimations of $\kappa$ and $D$ in all the analyses, as can clearly be seen in Figs.~\ref{fig:2:K_D_vs_Texpo_500hz}(a) and \ref{fig:2:K_D_vs_Texpo_500hz}(c). While $\delta$ increases, the values of $\kappa$ and $D$ linearly move away from the expected value: $\kappa$  is more and more overestimated, while $D$ is increasingly underestimated. An exception is the ACF method, which shows an opposite behavior and gives the best results under these conditions. These values were obtained from the fitting procedure ignoring the first data point of the ACF ($\rm{ACF}(\tau_{\ell=0})$), which, as we saw in previous section, carries most of the bias arising from the integration time. Moreover, in the MSD and in the ACF only the first data points corresponding at most to $6\,\tau_{\rm ot}$ ($\tau_\ell\leq6\,\tau_{\rm ot}$) were taken into consideration for the fitting.

We can see from the estimated values that the bias with the integration time gets less important as the sampling frequency raises, as shown in Figs.~\ref{fig:3:K_D_vs_Texpo_1500hz} and \ref{fig:4:K_D_vs_Texpo_3500hz}, since in the extreme case with $f_{\rm s}=3500\,\rm{Hz}$ and $\delta=150\,\mu \rm{s}$ the sampling frequency is at least one order of magnitude larger than the characteristic frequency of the trap and the integration time is one order of magnitude shorter than the characteristic time of the trap ($f_{\rm s} \gg f_{\rm{ot}}$ and $\delta \ll \tau_{\rm{ot}}$). The bias of $\kappa$ and $D$ with respect to $\delta$ and sampling frequency disappear for all the methods when the generalized formulas are used (see Figs.~\ref{fig:2:K_D_vs_Texpo_500hz}(b,d), \ref{fig:3:K_D_vs_Texpo_1500hz}(b,d), and \ref{fig:4:K_D_vs_Texpo_3500hz}(b,d)). Interestingly, all the diffusion constants estimated by the generalized analyses fall within the confidence interval of $D^*$.

All of the experimental data points shown in Figs.~\ref{fig:2:K_D_vs_Texpo_500hz}, \ref{fig:3:K_D_vs_Texpo_1500hz}, and \ref{fig:4:K_D_vs_Texpo_3500hz} were estimated with the same number of samples ($N_{\rm s}=10^5$).
This means that the data obtained for one experiment at $500\,\rm{Hz}$ lasted $3.3$ minutes, while those with $f_{\rm s}=3500\,\rm{Hz}$ lasted only $28.5\,\textrm{s}$.
To analyze the dependence of these results on the number of samples and on the total acquisition time, we show the convergence of the estimation of some data points in Figs.~\ref{fig:2:K_D_vs_Texpo_500hz}, \ref{fig:3:K_D_vs_Texpo_1500hz}, and \ref{fig:4:K_D_vs_Texpo_3500hz} as a function of time and number of samples, as it is explained in the following section.      

\section{Influence of the trajectory length}
\label{sec_trajectorylength}

\begin{figure}[H]
    \centering
    \medskip
     \includegraphics[width=13cm]{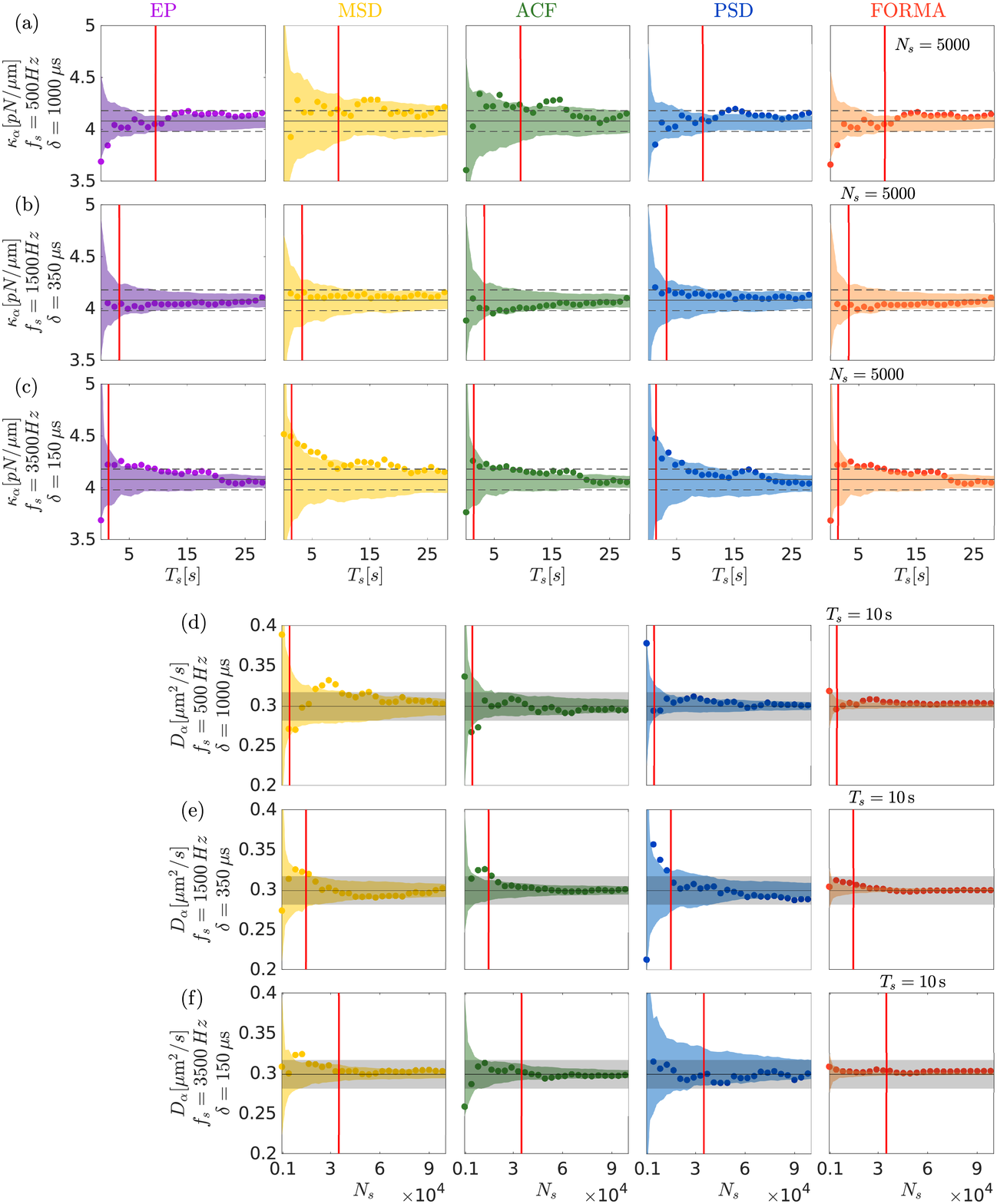}
    \caption{{\bf Performance of the generalized methods to estimate the stiffness and diffusion.} (a-c) The stiffness is estimated as a function of the total sampled time $T_{\rm s}$, while the diffusion (d-f) is estimated as a function of the number of samples $N_{\rm s}$ in the trajectory (at three sampling frequencies $f_{\rm s}$ and three long integration times $\delta$). Experimental results are shown with solid dots while colored shaded areas show confidence intervals estimated by Monte Carlo simulations considering $\bar{\kappa}_\alpha=4.08\,\rm{pN/ \mu m}$ and $\bar{D}_\alpha=0.299\,\rm{\mu m^2/s}$. Dashed lines in (a-c) depict the $10\%$ error range of $\bar{\kappa}$ and gray shaded areas in (d-f) depict the range of the expected value $D^*=0.295\pm0.02\,\rm{\mu m^2/s}$. The vertical red lines in (a-c) correspond to $N_{\rm s}=5000$ samples and the vertical red lines in (d-f) correspond to $T_{\rm s}=10\,\rm{s}$ acquisition time.}
    \label{fig:5:kvsTN}
\end{figure}

While it is intuitive to assume that both the total time $T_{\rm s}$ and the number of data points $N_{\rm s}$ have to be large in order to obtain accurate and precise values, the increase of these two parameters do not necessarily lead to better estimates, as we show in Fig.~\ref{fig:5:kvsTN}. At first glance, as Figs.~\ref{fig:5:kvsTN}(a-c) show, the convergence of the stiffness with respect to the total sampled time $T_{\rm s}$ does not show any significant differences for the three considered sampling frequencies. After a long enough sampled time of about $T_{\rm s}=10\,\rm{s}$, all the methods yield good results within an error of $10\%$ (dashed-black lines). Since these results are independent of the sampling frequency, for a given $T_{\rm s}$ the number of data points do not affect importantly the result. Fig.~\ref{fig:5:kvsTN}(a) has only $N_{\rm s}=14300$ samples, while Fig.~\ref{fig:5:kvsTN}(c) has $N_{\rm s}=10^5$ samples (see red vertical lines stressing the time that correspond to $N_{\rm s}=5000$ samples in the three cases). This result has important consequences if our main goal is to determine the stiffness, since an accurate and precise estimation will require enough time only, much larger than the characteristic relaxation time of the trap, and a faster camera or QPD will not necessarily improve the experimental accuracy. Moreover, this result could play a very important role when the experiment requires a long integration time, for instance when the light used for detection is very dimmed, whether it is for practical reasons or because the interesting sample could be photodamaged or altered with high intensity illumination.

On the contrary, a good estimation of the diffusion constant requires a large number of samples independently of the total time acquired. Figs.~\ref{fig:5:kvsTN}(d-f) exhibit the behavior of the diffusion with the number of samples $N_{\rm s}$ for all the methods, showing no dramatic dependence for each sampling frequency. In this case, we fixed the maximum number of samples to $N_{\rm s}=10^5$ corresponding to $T_{\rm s}=28.6\,\rm{s}$ for the highest sampling frequency (Fig.~\ref{fig:5:kvsTN}(f)) and to $T_{\rm s}=3.3\,\rm{min}$ for the lowest sampling frequency (Fig.~\ref{fig:5:kvsTN}(d)). Due to this property, the estimation of the diffusion constant is not limited by the total acquisition time, so it can be retrieved much faster than the stiffness by increasing the speed of the camera or detector. This is important in practice if the main goal is to estimate the diffusion, drag coefficient or the viscosity in conditions where these parameters are not stationary. For instance, using the highest sampling frequency in our experiments of $f_{\rm s}=3500\,\rm{Hz}$ (Fig.~\ref{fig:5:kvsTN}(f)), the time taken to have a good estimation, under the gray shaded area in Figs.~\ref{fig:5:kvsTN}(d-f), will require only a few seconds, or even a fraction of a second using FORMA.

Comparing the performance of the generalized methods, in general all of them have a very similar behavior when estimating the stiffness. Looking at the diffusion as a function of the number of samples $N_{\rm s}$ (Fig.~\ref{fig:5:kvsTN}(d-f)), the rhythm at which the various methods converge is very different. While the MSD, ACF, and PSD methods have an error of about $10\%$ at $N_{\rm s}=2\times10^4$ samples, FORMA achieves a much lower error, lower than $1\%$. The PSD seems to be more sensitive to sampling frequency, showing larger errors at the highest sampling frequencies. This is probably an error arising from a deficient representation of the PSD at low frequencies, biasing the fitting in this case, meaning that a longer total time could benefit this estimation. It is important to stress that these estimations were realized with standard non-linear fitting, but weighted non-linear fitting would improve the estimations making them closer to those obtained with FORMA, at the expense of using more computing time and more sophisticated algorithms. Moreover, the complete expression for the PSD shown in Eq.~\eqref{eq:PSDcomplete} may also contribute to improve these results. 

\section{Discussion}\label{sec_discussion}

Integration time, sampling frequency and data length of the trajectory of the particle in an OT affect the estimation of stiffness and diffusion constant in all the standard methods normally employed to extract these observables. 
All the standard methods enable accurate results when (i) the sampling frequency $f_{\rm{s}}$ is much higher than the characteristic frequency of the trap $f_{\rm{ot}}$, (ii) the integration time $\delta$ is much shorter than the characteristic relaxation time $\tau_{\rm{ot}}$, and (iii) the total acquisition time is several orders of magnitude larger than the characteristic relaxation time of the optical trap. 

When these ideal conditions start to weaken, the estimators for $\kappa$ and $D$ based on standard methods diverge from the real values.  First, thinking of an ideal experimental situation where the integration time is negligible and only the sampling frequency is reduced until a value close or even lower than the characteristic frequency of the trap, most of the methods seem to give reasonable predictions as long as the sampling frequency remains similar to the characteristic frequency. Under this condition, FORMA shows less reliable results, which is understandable since this method is based on the assumption of high sampling frequencies  \cite{Perez2018}. If the sampling frequency gets much lower than the characteristic frequency, then an important loss of information at short times in the MSD and ACF could bias the fitting of their respective analytical functions, affecting particularly the prediction of $D$. The predictions by means of the PSD may also be affected if the Nyquist frequency (corresponding to half the sampling frequency) does not reach values around the corner frequency in order to properly define the PSD and obtain a good fitting, enabling a good estimation of $\kappa$ and $D$. The potential and equipartition methods are probably the least affected under these circumstances as long as the sampled trajectory is long enough in order minimize the statistical error in the experimental histogram or the variance.

Second,  going a bit farther from the ideal conditions, as the integration time increases, the deviation of $\kappa$ and $D$ from their estimated values obtained using the standard methods increase as well. For the EP, MSD, and PSD the trend is more or less similar: the stiffness gets overestimated  and diffusion underestimated as the integration time increases, as can be seen in Figs.~\ref{fig:2:K_D_vs_Texpo_500hz}, \ref{fig:3:K_D_vs_Texpo_1500hz}, and \ref{fig:4:K_D_vs_Texpo_3500hz}.  As can be seen in Fig.~\ref{fig:1:estimators_ideal_vs_notideal}, the main deviation of the ACF from the theoretical value occurs at $\rm{ACF}(\tau_{\ell=0})$. Ignoring this value allows to get very good estimations of the stiffness and diffusion, even when the data set are far from the ideal conditions, as can be seen in Figs.~\ref{fig:2:K_D_vs_Texpo_500hz}, \ref{fig:3:K_D_vs_Texpo_1500hz}, and \ref{fig:4:K_D_vs_Texpo_3500hz}. On the opposite end, there is FORMA,  which is a reliable, simple and fast method when the sampling frequency is much higher than the characteristic time of the OT \cite{Perez2018}. In this respect, low sampling and non-zero integration time drastically degrade the estimations obtained by means this method in its standard form.

Incorporating both the integration time and frequency sampling into the estimators in all the common methods improve the estimates drastically, making them more accurate and precise. For a sampled trajectory with a large number of data taken over a long time, the estimations for all these generalized methods are comparable, showing a high degree of accuracy and precision, even if the integration time is large or the sampling frequency low, as can be seen in Figs.~\ref{fig:2:K_D_vs_Texpo_500hz}, \ref{fig:3:K_D_vs_Texpo_1500hz}, and \ref{fig:4:K_D_vs_Texpo_3500hz}, for the generalized formulas. At this level, the main differences in the methods will come from the way in which the experimental estimators are used or compared with model functions in order to extract $\kappa$ and $D$. MSD, ACF, and PSD require an intermediate routine to fit their experimental estimators, commonly using well-established non-linear fitting routines, or some intermediate procedures such as data selection or windowing in the case of the PSD, requiring in many cases preliminary knowledge of the system to optimize the solutions and computing time. In this respect, EP and FORMA contrast with the other methods in the sense that there is not any fitting involved in their solutions and it is not necessary to incorporate any extra information or any intermediate routine, making them very fast and reliable. This can clearly be seen in  Figs.~\ref{fig:2:K_D_vs_Texpo_500hz}, \ref{fig:3:K_D_vs_Texpo_1500hz}, and \ref{fig:4:K_D_vs_Texpo_3500hz}, where EP and FORMA in their generalized form, overall, retrieve the most accurate estimations of $\kappa$ and $D$. 

In general, the estimations are expected to improve as the amount of information contained  in the trajectories increases. This information is given by the number of data sampled and the total acquisition time. One might be tempted to increase them by recording longer times and/or by increasing the number of data points in the trajectory by means of the sampling frequency. However, one must make considerations about the specific requirements imposed by the experiment; for instance,  there are many situations where the sampling frequency can in principle be increased to very high values (e.g., using a fast camera system or a quadrant photodiode), but degrading the signal-to-noise ratio and hence the estimations of $\kappa$ and $D$. 

In practice, the total acquisition time cannot be extended indefinitely because the experimental conditions may be non-stationary or this would increase considerably the computational resources to record, store and analyze the data. In this respect, there are better approaches depending on whether one needs an accurate value of the stiffness $\kappa$ or of the diffusion constant  $D$.   For instance, if one wants to determine the stiffness with high accuracy, this will require enough acquisition time only, much larger than the characteristic time of the trap, while a faster camera or detector will not necessarily improve the estimation. On the other hand,  contrary to the stiffness, the diffusion constant can be retrieved much faster by increasing the speed of the camera or detector (Fig.~\ref{fig:5:kvsTN}). Additionally,  we can notice important discrepancies between the methods when insufficient information is contained in the trajectories.  Particularly, in Fig.~\ref{fig:5:kvsTN} we notice that the stiffness for all the methods behaves very similarly, but for the case of the diffusion constant,  FORMA  in its generalized form shows much better accuracy and precision than the other ones. 
 
The analysis shown in this work is directly applicable to data acquired by means of a standard digital camera, where the integration time is typically well defined. When dealing with quadrant photodiodes, the integration time may not be accessible, but a slightly adapted version of the analysis presented here may be still applicable. For example, the generalized MSD and ACF allow one to estimate $\delta$ in addition to $D$ and $\kappa$. Similar considerations apply for the cases where an effective integration time is indirectly implemented in the data processing by means of low-pass filtering.

\section{Conclusions}\label{sec_conclussions}

In this work, we have derived  generalized formulas for the most common OT calibration methods by incorporating the integration time and sampling frequency. The effects of these factors on the estimates of the stiffness and diffusion were analyzed theoretically, numerically, and experimentally.

We have shown that the integration time has important consequences for all the standard methods, leading to errors of up to $50\%$ in some extreme cases. The ACF seems to be an exception, behaving better than the other methods if the zero lag time ($\text{ACF}(\tau_{\ell=0})$) is ignored in the fitting procedure .
On the other hand, the generalized formulas give accurate and precise results independently of the technique, with errors that are less than $10\%$ for a few seconds, or a thousand of data sampled under our experimental conditions. The results can be drastically improved by incorporating either more data or sampling for a longer time, independently of the integration time or the sampling frequency. Particularly, we show that the generalized formulas give very good estimations of the stiffness if the sampled time is long enough, longer than an order of magnitude of the characteristic relaxation time, without worrying about the sampling frequency. In stark contrast, the diffusion constant can be estimated very fast and accurately with all generalized methods, even when the sampled time is very short or with very few data. More interestingly, the generalized expression for FORMA enables to retrieve very good estimations of the diffusion within an error of less than $10\%$ with some thousands, or even hundreds, of data points, making feasible to estimate the diffusion extremely fast with a high accuracy, depending mainly on the speed of the detector. In our case, we showed that a sampling frequency of $3500\,\rm{Hz}$ enables to obtain the diffusion within less than a second and with much less relative error than $10\%$ for a stiffness of $4.08\,\rm{p N/\mu m}$. 

This work paves the way to extend the applicability of common methods to contexts where the experimental conditions or the limitations of the setup do not allow measurements in ideal conditions, i.e., at short integration time, high frequency sampling and large amount of data, to obtain fast and reliable information about the stiffness and diffusion in  OT.

\begin{table}[H]
	\begin{center}
	\begin{scriptsize}
	\begin{tabular}{m{5cm} m{7cm}}
	\hline
 \multicolumn{2}{p{12cm}}{{\bf Equipartition (EP) analysis.} The stiffness $\kappa$ is computed from the the variance, $s^2_{N_{\rm s}}=\frac{1}{N_{\rm s}-1}\sum^{N_{\rm s}}_{n=1}( \tilde{x}_n-\tilde{x}_{\rm eq})^2$, with $\tilde{x}_{\rm eq}=\frac{1}{N_{\rm s}}\sum^{N_{\rm s}}_{n=1}\tilde{x}_n$, by means of the equipartition theorem:}\\

{\bf Standard}: 
 \setlength\abovedisplayskip{2pt}
  \setlength\belowdisplayskip{2pt}
\begin{equation*}  
   \kappa=\frac{k_{\rm B} T}{s^2_{N_{\rm s}}}\,. \tag{\ref{eq:convep}}
   \end{equation*}
   & 
   {\bf Generalized}: 
   \setlength\abovedisplayskip{2pt}
    \setlength\belowdisplayskip{2pt}
 \begin{equation*}
\kappa=\frac{k_B T}{s^2_{N_{\rm s}}}\mathcal{F}(\alpha)\,\quad \text{with} \quad \mathcal{F}(\alpha)=\frac{e^{-2\alpha}+2\alpha-1}{2\alpha^2}\,,\tag{\ref{eq:revep}}
 \end{equation*}

with $\alpha=\delta/2\tau_{\rm ot}$ and $\tau_{\rm ot}=\gamma/\kappa$. Since Eq.~\eqref{eq:revep}  depends on $\alpha$, we can proceed either assuming $\gamma$ known or perform some experiments with different values of $\delta$ in order to solve for $\kappa$ and $\tau_{\rm ot}$.
 \\
 
\hline
\multicolumn{2}{p{12cm}}{{\bf Mean squared displacement (MSD) analysis.}
 		The statistical estimator is computed by  $\text{MSD}(\tau_\ell)=\frac{1}{N_{\rm s}-\ell}\sum^{N_{\rm s}-\ell}_{n=1}(\tilde{x}_{n+\ell}-\tilde{x}_n)^2$. This dataset is fitted to the model function, with the free parameters $\kappa$ and $\tau_{{\rm ot}}$,}\\ 
 	{\bf Standard}:
   \setlength\abovedisplayskip{2pt}
    \setlength\belowdisplayskip{2pt}
\begin{equation*}
\text{MSD}(\tau)=\frac{2 k_{\rm B} T}{\kappa}\left(1-e^{-\tau/\tau_{{\rm ot}}}\right)\,. \tag{\ref{eq:convmsd}}
\end{equation*}
&
{\bf Generalized}:
 \setlength\abovedisplayskip{2pt}
 \setlength\belowdisplayskip{2pt}
\begin{equation*}
\text{MSD}(\tau_\ell)=\frac{2 k_B T}{\kappa}\left\{\mathcal{F}(\alpha)-\left[\frac{\sinh(\alpha)}{\alpha} \right]^2 e^{-|\tau_{\ell}|/\tau_{\text{ot}}}\right\}\,. \tag{\ref{eq:revmsd}}
\end{equation*}

Alternatively, we could infer the three free parameters 

$a=2 k_B T \mathcal{F}(\alpha)/\kappa$, $b=2 k_B T (\sinh(\alpha)/\alpha)^2/\kappa$ and $c=\tau_{\rm ot}$, from which, we could infer $\tau_{\rm ot}$ and $\kappa$ without knowing $\delta$, which actually could be inferred as well.
\\
\hline
\multicolumn{2}{p{12cm}}{		{\bf  Autocorrelation function (ACF) analysis.} 
  			The statistical estimator is $\text{ACF}(\tau_\ell)=\frac{1}{N_{\rm s}-\ell}\sum^{N_{\rm s}-\ell}_{n=1}(\tilde{x}_{n+\ell}\tilde{x}_n)$. These data points are fitted to the model function, with the free parameters $\kappa$ and $\tau_{{\rm ot}}$,}\\
  			{\bf Standard}:
      \setlength\abovedisplayskip{2pt}
       \setlength\belowdisplayskip{2pt}
 		\begin{equation*}
        \text{ACF}(\tau)=\frac{k_{\rm B} T}{\kappa} e^{-\tau/\tau_{{\rm ot}}}\,,\tag{\ref{eq:convacf}}
        \end{equation*}
 &
 {\bf Generalized}: 
  \setlength\abovedisplayskip{0pt}
   \setlength\belowdisplayskip{0pt}
 \begin{equation*}
\text{ACF}(t_n-t_m)=\frac{k_B T}{\kappa}\left\{
\begin{array}{ll}
\left[\frac{\sinh(\alpha)}{\alpha} \right]^2  e^{-|t_n-t_m|/\tau_{\text{ot}}}& t_n\neq t_m\,,\\
\mathcal{F}(\alpha)& t_n= t_m\,.
\end{array}\right.\,.\tag{\ref{eq:revacf}}
\end{equation*}
If we know $\delta$ we could use the solution for $t_n\neq t_m$ and solve for $\kappa$ and $\tau_{\rm ot}$. Alternatively, we could divide the inference problems into two parts corresponding to zero and non-zero lag time and get the values of $\tau_{\rm ot}$, $\kappa$ and $\alpha$. \\
 	\hline
 \multicolumn{2}{p{12cm}}{{\bf Power spectral density (PSD) analysis.} The expected values of the aliased PSD are computed by means of the Fast Fourier Transform,
   \setlength\abovedisplayskip{2pt}
  \setlength\belowdisplayskip{2pt}
 \begin{equation*}
P(f_k)=\frac{\left<|\hat{x}(f_k)|^2\right>}{T_{\rm s}}
	=
	\frac{\Delta t^2}{T_{\rm s}}
	\left<|\rm{FFT}\{\{\tilde{x}_j\}^{N_{\rm s}}_{j=1}\}_k|^2\right> 
	\ \ \ 
	\mbox{with} 
	\ \ \ 
	f_k
	=
	\frac{k}{T_{\rm s}} 
	\ \ \ \
	\mbox{for} 
	\ 
	k=0,1,2,\ldots,(N_{\rm s}-1)/2\,,
 \end{equation*}
 where the notation $\bracket{\cdots}$ indicates the expected value estimated either by the average of many experimental replicas or through data compression. These values are then fitted to the model function with $f_{\rm c}=\kappa/2\pi\gamma$ and $D$ as free parameters,}
 \\
 {\bf Standard}: 
  \setlength\abovedisplayskip{2pt}
   \setlength\belowdisplayskip{2pt}
\begin{equation}
P(f)
	=
		\frac{
		{\Delta \tilde{x}}^2/f_{\rm s}
	}{
		1+c^2-2c\cos(2\pi f/f_{\rm s})
	},\tag{\ref{eq:convpsdal}}
\end{equation}
where $\Delta \tilde{x}=((1-c^2)D/2\pi f_{{\rm c}})^{1/2}$ and $c=e^{-2\pi f_{{\rm c}}/f_{\rm s}}$. 

&
{\bf Generalized}: 
 \setlength\abovedisplayskip{2pt}
  \setlength\belowdisplayskip{2pt}
\begin{multline*}
P(f)=D\tau_{\text{ot}}\Delta t\left(\left[\frac{\sinh(\alpha)}{\alpha} \right]^2\frac{\sinh(\Delta t/\tau_{\text{ot}})}{\cosh(\Delta t/\tau_{\text{ot}})-\cos(2\pi f\Delta t)}\right.\\
\left.+\frac{\alpha-\cosh(\alpha)\sinh(\alpha)}{\alpha^2}\right)+\mathcal{O}(N_{\rm s}^{-1})\,.\tag{\ref{eq:revpsdal}}
\end{multline*}\\ 	
 \hline
 \multicolumn{2}{p{12cm}}{{\bf Force reconstruction via maximum-likelihood-estimator (FORMA) analysis.} The stiffness $\kappa$ and diffusion $D$ (or $\gamma$)  are directly computed from $\mathcal{D}$ via the formulas}\\
 
 {\bf Standard}:
  \setlength\abovedisplayskip{2pt}
   \setlength\belowdisplayskip{2pt}
\begin{equation}
\begin{split}
	\frac{\kappa}{\gamma}
	=
	-\frac{
		\sum_{n=1}^{N_{\rm s}-1}
			\tilde{x}_{n}
			\frac{
				\tilde{x}_{n+1} -\tilde{x}_{n}
			}{
				\Delta t
			} 
	}{
		\sum_{n=1}^{N_{\rm s}-1} [\tilde{x}_{n}]^2
	},
\\	
	D
	=
	\frac{\Delta t}{2 (N_{\rm s}-1)}
	\sum_{n=1}^{N_{\rm s}-1}
		\left(
			\frac{
				\tilde{x}_{n+1}-{\tilde{x}_{n}}
			}{
				\Delta t
			}
			+
			\frac{
				\kappa
			}{
				\gamma
			} 
			\tilde{x}_{n}
		\right)^2\,. 
		\end{split} \tag{\ref{eq:convforma}}
\end{equation}

  		&
{\bf Generalized}:
 \setlength\abovedisplayskip{2pt}
  \setlength\belowdisplayskip{2pt}
\begin{equation*} 
\begin{split}
\mathcal{G}(\alpha) e^{-\Delta t/\tau_{\rm{ot}}}=\frac{\mathcal{T}_2}{\mathcal{T}_3}\,,
\\
\quad\quad\frac{k_{B}T}{\kappa} \mathcal{F}(\alpha)=\frac{\mathcal{T}_1-\frac{\mathcal{T}_2^2}{\mathcal{T}_3}}{1-\left(\frac{\mathcal{T}_2}{\mathcal{T}_3}\right)^2}\,,
\end{split}\tag{\ref{eq:revforma}}
\end{equation*}
with $\mathcal{G}(\alpha)=\left[\frac{\sinh(\alpha)}{\alpha}\right]^2/\mathcal{F}(\alpha)\, $, $\mathcal{T}_1=\frac{1}{N_{\rm s}-1}\sum_{n=1}^{N_{\rm s}-1}[\tilde{x}_{n+1}]^2\,,\quad    \quad $
$\mathcal{T}_2=\frac{1}{N_{\rm s}-1}\sum_{n=1}^{N_{\rm s}-1}\tilde{x}_{n+1}\tilde{x}_{n}
\ \ \ \ \text{and} \quad \mathcal{T}_3=\frac{1}{N_{\rm s}-1}\sum_{n=1}^{N_{\rm s}-1}[\tilde{x}_{n}]^2.$\\
\hline
 	\end{tabular}
	\end{scriptsize}
\caption{\textbf{Overview of the calibration analyses in the overdamped regime}. These methods allow to estimate the stiffness $\kappa$ of the optical trap and, in most of the cases, the diffusion constant $D$ of the optically trapped particle from the sampled trajectory of the  particle $\mathcal{D}\equiv\{ \tilde{x}_{j}\}_{j=1}^{N_{\rm s}}$ at sampled time $\Delta t$ and integration time $\delta$. The standard model functions assume much shorter sampling and integration times than the characteristic time of the optical trap. On the other hand, the generalized model functions take into account arbitrary sampling frequency and integration time, allowing to drastically improve the calibration of OT even if data acquisition is very far from the ideal conditions.}
	\label{tab:methods}
	\end{center}
\end{table}

\begin{appendices}

\section{Brownian dynamics simulations}
\label{sec:simulations}
The dynamics of the a particle in an optical trap is described by the Langevin equation defined in Eq.~\eqref{eq:langevin}, which can be solved by means of the Wiener process as follows \cite{volpe2013simulation,jones_marago_volpe_2015}:
\begin{equation}\label{Wienner}
    x_{k+1}=x_k-\frac{\kappa}{\gamma}x_k\Delta t_{\rm s}+\sqrt{D\Delta t_{\rm s}}w_k,
\end{equation}
where $w_k$ is a normally distributed random variable with zero mean and unitary variance. The trajectory $\{x_{k}\}^{M}_{k=1}$ is generated with a fixed time interval between successive points of $\Delta t_{\rm s}=1\,\rm{\mu s}$. To emulate the effect of a detection system with a given sampling frequency $f_{\rm s}$, it will suffice to select sub-trajectories $\{x_n\}_{n=1}^{N_{\rm s}}$ such that their consecutive points have a time difference of $\Delta t=1/f_{\rm s}$ larger tan $\Delta t_{\rm s}$. For the effect of a finite integration time  $\delta > 0$, one needs to find the closest integer of the ratio $ \delta/ \Delta t_{\rm s}$ and perform the average over the corresponding points in the trajectory  $\{x_k\}$ to obtain the sampled trajectory that corresponds to the one of a given sampling frequency and a finite integration time. In our simulations, we used $\kappa = 4.08\,\rm{pN/\mu m}$ and $D= 0.299\,\rm{\mu m^2/s}$. 

\section{Experimental methods}
\label{sec:experiments}
The OT was generated by means of a Gaussian laser beam (Coherent Verdi V6, $\lambda=532\,\rm{nm}$): this laser beam was expanded, phase corrected by a phase-only spatial light modulator (Hamamatsu X10468-04), relayed by a 1:1 telescope, and tightly focused by a water immersion objective (Olympus UPLANSApo $60\times$ with $\rm{NA}=1.2$), within an inverted microcope. The sample consists of spherical particles (Bangs Laboratories Inc., non-functionalized silica particles of diameter $d_{\rm p}=1.54\pm 0.10\,\rm{\mu m}$) suspended in an aqueous solution at laboratory temperature ($T=22\,\pm0.5^\circ\rm{C}$) and put inside a  sealed sample cell made with two coverslips and a spacer of about $100\,\mu\rm{m}$ thickness. The beam was focused deep into the sample cell ensuring that the 3D trapped particle was far from both the bottom and the upper coverslips (at least $10\,\rm{\mu m}$ from the bottom one), preventing in this way any significant hydrodynamic interactions with the walls of the sample cell. To perform long-term experiments with this microscope objective, which works with a thin layer of water between the lens and the bottom coverslip, we used a water dispenser based on capillary action to compensate for water loss that is constantly evaporating (see Ref.~\cite{Arzola2019} for more details). The dynamics of the particle was recorded with a CMOS camera (Basler Ace acA640-750um) at different sampling frequencies and integration times (see main text), and its position was obtained using standard video microscopy techniques \cite{crocker1996methods, Shattuck2017handbook}, with an accuracy of less than $5\,\rm{nm}$ in the position detection. 

\section{Derivation of the joint PDF of the corrected particle's trajectory}
\label{app:A}
Starting from the definition of the time-average position given in Eq.~\eqref{eq:def} and using the formal solution given by Eq. \eqref{eq:exactsolution}, we need to evaluate two terms. The first term reads:
\beeq{
\frac{1}{\delta}\int_{t_n-\delta/2}^{t_n+\delta/2} dt x_0 e^{- t/\tau_{\text{ot}}}&= x_0 \frac{e^{- t_n/\tau_{\text{ot}}}}{2\alpha} \left(e^{\alpha}-e^{-\alpha}\right)=\tilde{x}_0 e^{- t_n/\tau_{\text{ot}}}.
}
For the second term, the idea is to change the order of integration to express it as an integral over the whole trajectory of a function coupled with the noise. Thus, we write
\beeq{
 \frac{\sqrt{2D}}{\delta}\int_{t_n-\delta/2}^{t_n+\delta/2} dt\int_0^{t} ds W_x(s) e^{-(t-s)/\tau_{\text{ot}}} &= \frac{\sqrt{2D}}{\delta}\Bigg[\int^{t_n-\delta/2}_0 ds W_x(s)\int_{t_n-\delta/2}^{t_n+\delta/2} dte^{-(t-s)/\tau_{\text{ot}}} \\
& +\int^{t_n+\delta/2}_{t_n-\delta/2} ds W_x(s) \int_{s}^{t_n+\delta/2} dte^{-(t-s)/\tau_{\text{ot}}} \Bigg]\\
&=\sqrt{2D}\int_0^T ds  W_x(s) g_n(s)\,,
}
where we have defined
\beeq{
g_n(s)&\equiv \Theta(t_{n}-\delta/2 -s)\frac{1}{\delta}\int_{t_n-\delta/2}^{t_n+\delta/2} dte^{-(t-s)/\tau_{\text{ot}}}\\
&+\Theta(s-t_{n}+\delta/2)\Theta(t_{n}+\delta/2- s) \frac{1}{\delta}\int_{s}^{t_n+\delta/2} dte^{-(t-s)/\tau_{\text{ot}}}\,,
}
being $\Theta(x)$ the Heaviside step function. The two integrals appearing in the above definition of $g_n(s)$ can be easily done, viz.
\begin{eqnarray}
\frac{1}{\delta}\int_{t_n-\delta/2}^{t_n+\delta/2} dte^{-(t-s)/\tau_{\text{ot}}}&=&\frac{1}{2\alpha}\left[e^{-(t_n-\delta/2-s)/\tau_{\text{ot}}}-e^{-(t_n+\delta/2-s)/\tau_{\text{ot}}}\right]\\
&=&\frac{\sinh(\alpha)}{\alpha}e^{-(t_n-s)/\tau_{\text{ot}}}\,,\\
\frac{1}{\delta}\int_{s}^{t_n+\delta/2} dte^{-(t-s)/\tau_{\text{ot}}}&=&\frac{1}{2\alpha}\left[1-e^{-(t_n+\delta/2-s)/\tau_{\text{ot}}}\right]\,.
\end{eqnarray}
We finally obtain the following expression for the stochastic dynamics of the corrected trajectory:
\beeq{
\tilde{x}(t_n)=\tilde{x}_0 e^{- t_n/\tau_{\text{ot}}}+\sqrt{2D}\int_0^T ds  W_x(s)g_n(s)
}
with
\beeq{
g_n(s)&\equiv \Theta(t_{n}-\delta/2 -s)\frac{\sinh(\alpha)}{\alpha}e^{-(t_n-s)/\tau_{\text{ot}}}\\
&+\Theta(s-t_{n}+\delta/2)\Theta(t_{n}+\delta/2- s) \frac{1}{2\alpha}\left[1-e^{-(t_n+\delta/2-s)/\tau_{\text{ot}}}\right].
\label{eq:gn}
}
Now, we are in position to derive the joint PDF, denoted here $p\{\tilde{x}_n\}_{n=1}^{N_{\rm s}}$,  of observing the corrected particle's trajectories at the blurry points $\tilde{x}_1,\ldots, \tilde{x}_{N_{\rm s}}$ and times $t_1,\ldots, t_{N_{\rm s}}$. Mathematically, this is given by:
\beeq{
p(\{\tilde{x}_n\}_{n=1}^{N_{\rm s}})=\bracket{\prod_{n=1}^L\delta \left(\tilde{x}_n-\tilde{x}(t_n)\right)}_{W_x}\,,
}
where $\bracket{(\cdots)}_{W_x}$ denotes the average over all possible realization of the noise's trajectory $W_x(t)$, whose probability measure is given by the following path integral:
\beeq{
\mathcal{P}[W_x(t)]=\frac{1}{Z}\exp\left[-\frac{1}{2}\int_0^T ds W_x^2(s)\right]\,,
}
where $Z$ is a normalization constant. Let us now derive an exact expression for the joint PDF. To do this, we write
\beeq{
\bracket{\prod_{n=1}^{N_{\rm s}}\delta \left(\tilde{x}_n-\tilde{X}(t_n)\right)}_{\xi}=\bracket{\int\left[\prod_{n=1}^{N_{\rm s}}\frac{dy_n}{2\pi}\right]\exp\left[i\sum_{n=1}^{N_{\rm s}} y_n \left(\tilde{x}_n-\tilde{X}(t_n)\right)\right]}_{W_x}\\
=\int\left[\prod_{n=1}^{N_{\rm s}}\frac{dy_n}{2\pi}\right]\exp\left(i\sum_{n=1}^{N_{\rm s}} y_n \tilde{x}_n\right)\bracket{\exp\left[-i\sum_{n=1}^{N_{\rm s}} y_n \tilde{X}(t_n)\right]}_{W_x}\,.
}
The average over all possible realizations of the noise $W_x(t)$ has the following form:
\beeq{
\bracket{\exp\left[-i\sum_{n=1}^{N_{\rm s}} y_n \tilde{X}(t_n)\right]}_{W_x}&=\frac{1}{Z}\int \mathcal{D}W_x(t) \exp\Bigg[-\frac{1}{2}\int_0^Tds W_x^{2}(s)\\
&-i\sum_{n=1}^{N_{\rm s}} y_n \left(\tilde{x}_o e^{- t_n/\tau_{\text{ot}}} +\sqrt{2D}\int_0^Tds W_x(s) g_n(s)\right)\Bigg]\\
&=e^{-i\sum_{n=1}^{N_{\rm s}} y_n \tilde{x}_0 e^{- t_n/\tau_{\text{ot}}}}\\
&\times\frac{1}{Z}\int \mathcal{D}W_x(t) \exp\Bigg[-\frac{1}{2}\int_0^Tds W_x^{2}(s)\\
&-i\sqrt{2D}\int_0^Tds W_x(s)\sum_{n=1}^{N_{\rm s}} y_n g_n(s)\Bigg]\,.
}
But the path integral over $W_x(t)$ is easy to evaluate yielding:
\beeq{
&\frac{1}{Z}\int \mathcal{D}W_x(t) \exp\left[-\frac{1}{2}\int_0^Tds W_x^{2}(s)-i\sqrt{2D}\int_0^Tds W_x(s)\sum_{n=1}^{N_{\rm s}} y_n g_n(s)\right]\\
&=\frac{1}{Z}\int \mathcal{D}W_x(t)\times \\ & \ \ \ \ \ \ \ \ \ \exp\left[-\frac{1}{2}\int_0^Tds \left(W_x(s)+i\sqrt{2D}\sum_{n=1}^{N_{\rm s}} y_n g_n(s)\right)^2-D\sum_{n,m=1}^{N_{\rm s}} y_ny_m \int_0^Tdsg_n(s)g_{m}(s)\right]\\
&=\exp\left[-\frac{1}{2}\sum_{n,m=1}^{N_{\rm s}} y_n I_{nm}y_m\right]\,,
}
where we have defined
\beeq{
I_{nm}&=2D\int_0^Tdsg_n(s)g_{m}(s)\,.
\label{eq:covariant}
}
Gathering the results, we have that:
\beeq{
p(\{\tilde{x}_n\}_{n=1}^{N_{\rm s}})=\int\left[\prod_{n=1}^{N_{\rm s}}\frac{dy_n}{2\pi}\right]\exp\left(i\sum_{n=1}^{N_{\rm s}}y_n\left( x_n-\tilde{x}_0 e^{- t_n/\tau_{\text{ot}}}\right)-\frac{1}{2}\sum_{n,m=1}^{N_{\rm s}} y_n I_{nm}y_m \right)\,.
}
With respect to the auxiliary variables $y_n$ this is a multi-variate Gaussian integral, easy to carry out. The final result for the joint PDF of observing a blurry trajectory is:
\beeq{
p(\{\tilde{x}_n\}_{n=1}^{N_{\rm s}})&=\frac{1}{\sqrt{(2\pi)^{N_{\rm s}} \det I}}\exp\left[-\frac{1}{2}\sum_{n,m=1}^{N_{\rm s}} \left(x_n- \tilde{x}_0 e^{- t_n/\tau_{\text{ot}}}\right)[I^{-1}]_{nm}\left(x_m- \tilde{x}_0 e^{- t_m/\tau_{\text{ot}}}\right) \right],
}
where the matrix elements $I_{nm}$ are given by Eq. \eqref{eq:covariant}.
\section{Exact expression for the corrected covariant matrix $I$}
\label{app:B}
We now proceed to derive a simple expression for the corrected covariance matrix, introduced in Eq. \eqref{eq:covariant}, where the function $g_n(s)$ is defined in Eq. \eqref{eq:gn}. We first write the covariance matrix as the sum of three contributions, that is $I_{nm}=I^{(2)}_{nm}+I^{(1)}_{nm}+I^{(3)}_{nm}$, where
\beeq{
I^{(1)}_{nm}/2D&=\left[\frac{\sinh(\alpha)}{\alpha} \right]^2\int_0^Tds\Theta(t_n-\delta/2-s)\Theta(t_m-\delta/2-s)e^{-( t_n+t_m-2s)/\tau_{\text{ot}}}\\
I^{(2)}_{nm}/2D&=\left[\frac{\sinh(\alpha)}{\alpha} \right]\Bigg\{\int_0^Tds\Theta(t_n-\delta/2-s)\Theta(t_m+\delta/2-s)\Theta(s-t_m+\delta/2)\\
&\times\frac{e^{-( t_n-s)/\tau_{\text{ot}}}-e^{-(t_m+t_n+\delta/2-2s)/\tau_{\text{ot}}}}{2\alpha}\\
&+\int_0^Tds\Theta(t_m-\delta/2-s)\Theta(t_n+\delta/2-s)\Theta(s-t_n+\delta/2)\\
&\times\frac{e^{-( t_m-s)/\tau_{\text{ot}}}-e^{-(t_m+t_n+\delta/2-2s)/\tau_{\text{ot}}}}{2\alpha}\Bigg\},\\
I^{(3)}_{nm}/2D&=\int_0^Tds\Theta(t_n+\delta/2-s)\Theta(s-t_n+\delta/2)\Theta(t_m+\delta/2-s)\Theta(s-t_m+\delta/2)\\
&\times\frac{1-e^{-(t_n+\delta/2-s)/\tau_{\text{ot}}}}{2\alpha}\frac{1-e^{-(t_m+\delta/2-s)/\tau_{\text{ot}}}}{2\alpha}.
}
The evaluation of these integrals is relatively simple. Let us evaluate each carefully. For the integral $I^{(1)}_{nm}$, we notice that the two Heaviside functions indicate that the time integral over the $s$-variable must be carried out in the interval $s\in [0,\min(t_n,t_m)-\delta/2]$. Thus, we write
\beeq{
I^{(1)}_{nm}/2D&=\left[\frac{\sinh(\alpha)}{\alpha} \right]^2\int_0^Tds\Theta(t_n-\delta/2-s)\Theta(t_m-\delta/2-s)e^{-( t_n+t_m-2s)/\tau_{\text{ot}}}\\
&=\left[\frac{\sinh(\alpha)}{\alpha} \right]^2\int_0^{\min(t_n,t_m)-\delta/2}dse^{-( t_n+t_m-2s)/\tau_{\text{ot}}}\\
&=\left[\frac{\sinh(\alpha)}{\alpha} \right]^2 \frac{\tau_{\text{ot}}}{2}\left[e^{-( t_n+t_m-2(\min(t_n,t_m)-\delta/2))/\tau_{\text{ot}}}-e^{-( t_n+t_m)/\tau_{\text{ot}}}\right]\,.
}
Now, using the following expressions for the minimum and maximum of two real numbers, viz.
\beeq{
\min(t_n,t_m)=\frac{t_n+t_m}{2}-\frac{|t_n-t_m|}{2}\,,\quad\quad \max(t_n,t_m)=\frac{t_n+t_m}{2}+\frac{|t_n-t_m|}{2}\,,
}
we finally arrive at:
\beeq{
I^{(1)}_{nm}&= D\tau_{\text{ot}}\left[\frac{\sinh(\alpha)}{\alpha} \right]^2 \left[e^{-(|t_n-t_m|+\delta)/\tau_{\text{ot}}}-e^{-( t_n+t_m)/\tau_{\text{ot}}}\right]\,.
}
To carry out the integral $I^{(2)}_{nm}$, we will assume that  $|t_n-t_m|> \delta$, that is, the intervals  $[t_n-\delta/2,t_n+\delta/2]$ and $[t_m-\delta/2,t_m+\delta/2]$ never overlap (even for consecutive ones) unless $n=m$. Next, looking at the expression of $I^{(2)}_{nm}$, we see that the three Heaviside functions indicate that for the integral to be non-zero we must have that the intersections of the intervals $[0,t_n-\delta/2]\cap [t_m-\delta/2,t_m+\delta/2]$ and $[0,t_m-\delta/2]\cap [t_n-\delta/2,t_n+\delta/2]$ must be non-zero. This automatically implies that whenever $t_n=t_m$, the integral is zero. We are left to consider the two cases $t_m>t_n$ or $t_m<t_n$. This automatically implies that:
\beeq{
I^{(2)}_{nm}/2D&=\left(1-\delta_{nm}\right)\left[\frac{\sinh(\alpha)}{\alpha} \right]\Bigg[\Theta(t_n>t_m)\int_{t_m-\delta}^{t_m+\delta} ds\frac{e^{-( t_n-s)/\tau_{\text{ot}}}-e^{-(t_m+t_n+\delta/2-2s)/\tau_{\text{ot}}}}{2\alpha}\\
&+\Theta(t_m>t_n)\int_{t_n-\delta}^{t_n+\delta}ds\frac{e^{-( t_m-s)/\tau_{\text{ot}}}-e^{-(t_m+t_n+\delta/2-2s)/\tau_{\text{ot}}}}{2\alpha}\Bigg]\,.
}
Doing the integrals and rearranging terms, we finally arrive at the following expression
\beeq{
I^{(2)}_{nm}&= 2D\tau_{\text{ot}}\left(1-\delta_{nm}\right) \alpha\left[\frac{\sinh(\alpha)}{\alpha} \right]^3 e^{-\alpha-|t_m-t_n|/\tau_{\text{ot}}}\,.
}

Finally, looking at the expression of the integral $I^{(3)}_{nm}$, we observe that the integration interval is given by I$[t_n-\delta/2,t_n+\delta/2]\cap[t_m-\delta/2,t_m+\delta/2]$, so unless $t_n=t_m$ the integral is zero as the intervals do not overlap. Thus,
\beeq{
I^{(3)}_{nm}/2D&=\delta_{nm}\int_{t_n-\delta/2}^{t_n+\delta/2}ds\left(\frac{1-e^{-(t_n+\delta/2-s)/\tau_{\text{ot}}}}{2\alpha}\right)^2=\delta_{nm} \delta/2 \frac{4e^{-2\alpha }+4\alpha-3-e^{-4\alpha}}{8\alpha^3}\,,
}
and therefore
\beeq{
I^{(3)}_{nm}= 2D \tau_{\text{ot}}\delta_{nm} \alpha \frac{4e^{-2\alpha }+4\alpha-3-e^{-4\alpha}}{8\alpha^3}\,.
}
We can now add up the three contributions to obtain the following expression for $I_{nm}$, viz.
\beeq{
I_{nm}&= D\tau_{\text{ot}}\left[\frac{\sinh(\alpha)}{\alpha} \right]^2 \left[e^{-2\alpha-|t_n-t_m|/\tau_{\text{ot}}}-e^{-( t_n+t_m)/\tau_{\text{ot}}}\right]+2D\tau_{\text{ot}} \alpha\left[\frac{\sinh(\alpha)}{\alpha} \right]^3 e^{-\alpha-|t_m-t_n|/\tau_{\text{ot}}}\\
&+  2D \tau_{\text{ot}}\delta_{nm}\Bigg[ \alpha \frac{4e^{-2\alpha }+4\alpha-3-e^{-4\alpha}}{8\alpha^3}- \alpha\left[\frac{\sinh(\alpha)}{\alpha} \right]^3 e^{-\alpha}\Bigg].
}
Finally, using the following identities
\beeq{
\left[\frac{\sinh(\alpha)}{\alpha} \right]^2 e^{-2\alpha} +2\alpha \left[\frac{\sinh(\alpha)}{\alpha} \right]^3 e^{-\alpha}&=\left[\frac{\sinh(\alpha)}{\alpha} \right]^2\,,\\
\alpha \frac{4e^{-2\alpha }+4\alpha-3-e^{-4\alpha}}{8\alpha^3}- \alpha\left[\frac{\sinh(\alpha)}{\alpha} \right]^3 e^{-\alpha}&=\frac{\alpha-\cosh(\alpha)\sinh(\alpha)}{2\alpha^2}\,,
}
allow us to simplify the covariance matrix, yielding:
\beeq{
I_{nm}&= D\tau_{\text{ot}}\left[\frac{\sinh(\alpha)}{\alpha} \right]^2 \left[e^{-|t_n-t_m|/\tau_{\text{ot},x}}-e^{-( t_n+t_m)/\tau_{\text{ot},x}}\right]+  D \tau_{\text{ot}}\delta_{nm}\frac{\alpha-\cosh(\alpha)\sinh(\alpha)}{\alpha^2}
}.
\end{appendices}

\section*{Competing interests}
  The authors declare that they have no competing interests.

\section*{Author's contributions}
    A.V.A. had the original idea. G.V., I.P.C and A.V.A. supervised the study. I.P.C performed all theoretical derivations. L.P.G., M.S., A.C., G.P. and A.V.A. performed the experiments. L.P.G. and A.V.A. implemented the models, performed the simulations (some of which were initially carried out by I.P.C), did the data analysis and made the figures. All the authors discussed the results and L.P., I.P.C and A.V.A. wrote the draft of the article. All authors revised the final version of the article.
\section*{Acknowledgements}
We thank Agnese Callegari for her insights about the presentation of the results. 

\section*{Funding}
This work was supported by UNAM-PAPIIT IN111919, the H2020 European Research Council (ERC) Starting Grant ComplexSwimmers (grant number 677511), the Horizon Europe ERC Consolidator Grant MAPEI (grant number 101001267), and the Knut and Alice Wallenberg Foundation (grant number 2019.0079, received by G.V.).

\section*{Data availability}
Data underlying the results presented in this paper are not publicly available at this time but may be obtained from the authors upon reasonable request.

\bibliography{biblio}
\end{document}